\documentclass[a4paper,11pt,dvips]{article}
\usepackage{amssymb}
\usepackage{amsmath}
\usepackage{amscd}
\usepackage{latexsym}
\usepackage{epsfig}
\usepackage{multirow}
\usepackage{color}
\usepackage[bf,footnotesize]{caption}
\usepackage{subfigure}

\usepackage[dvips,colorlinks=true,a4paper=true,backref=false, linktocpage=true,
citecolor=blue,urlcolor=blue,linkcolor=blue,pdfpagemode=UseOutlines]{hyperref}

\hypersetup{%
  bookmarksnumbered=true,
  pdftitle = {HMC algorithm with multiple time scale
    integration and mass preconditioning},
  pdfsubject = {HMC algorithm for lattice QCD },
  pdfauthor = {C.Urbach, K.Jansen, A.Shindler, U.Wenger},
  pdfkeywords = {HMC, Hybrid Monte Carlo,
    Multiple time scale integration, mass preconditioning}
}
\usepackage{algorithm,algorithmic}

\newcommand{\Qpm}{Q^{\pm}}
\newcommand{\Qp}{Q^{+}}
\newcommand{\Qm}{Q^{-}}

\def\lsi{\raise0.3ex\hbox{$<$\kern-0.75em\raise-1.1ex\hbox{$\sim$}}}

\newcommand{\sign}{\operatorname{sign}}

\title{
  {\vspace{-3cm} \normalsize
    \hfill \parbox{40mm}{DESY 06-163\\
      LTH-716\\
      RM3-TH/06-14\\
      SFB/CPP-06-43\\
      WUB 06-04}}\\[10mm]
  Iterative methods for overlap and twisted mass fermions
}
\date{}

\author{T.~Chiarappa$^{\, 1}$, 
K.~Jansen$^{\, 2}$, 
K.-I.~Nagai$^{\, 3}$, 
M.~Papinutto$^{\, 4}$, \\
L.~Scorzato$^{\, 5}$, 
A.~Shindler$^{\, 2}$, 
C.~Urbach$^{\, 6}$, 
U.~Wenger$^{\, 7}$, 
I.~Wetzorke$^{\, 2}$\\ \ \\
{\small $^{1}$ Universit\`a Milano Bicocca,
   Piazza della Scienza 3, } \\
  {\small    I--20126 Milano, Italy} \\ \ \\
{\small $^{2}$ John von Neumann-Institut f{\"u}r Computing NIC,} \\
  {\small Platanenallee 6, D--15738 Zeuthen, Germany} \\ \ \\
{\small $^{3}$ Department of Physics, University of Wuppertal,} \\
  {\small Gaussstrasse 20, D--42119 Wuppertal, Germany} \\ \ \\
{\small $^{4}$ INFN Sezione di Roma Tre, } \\
  {\small Via della Vasca Navale 84, I--00146, Rome, Italy} \\ \ \\
{\small $^{5}$ ECT*, Strada delle Tabarelle 286,} \\
  {\small I--38050 Villazzano (TN), Italy} \\ \ \\
{\small $^{6}$  Division of Theoretical Physics, Department of Mathematical Sciences,} \\
  {\small The University of Liverpool, Liverpool L69 3BX, U.K.} \\ \ \\
  {\small $^{7}$ Institut f{\"u}r Theoretische Physik, } \\
  {\small ETH Z{\"u}rich, CH--8093
    Z\"urich, Switzerland} \\
}

\begin{document}
\maketitle
\thispagestyle{empty}
\begin{abstract}
  We present a comparison of a number of iterative solvers of linear systems
  of equations for obtaining the fermion propagator in lattice QCD. In
  particular, we consider chirally invariant overlap and chirally improved
  Wilson (maximally) twisted mass fermions. The comparison of both
  formulations of lattice QCD is performed at four fixed values of the pion
  mass between 230MeV and 720MeV.  For overlap fermions we address adaptive
  precision and low mode preconditioning while for twisted mass fermions we
  discuss even/odd preconditioning. Taking the best available algorithms in
  each case we find that calculations with the overlap operator are by a
  factor of 30-120 more expensive than with the twisted mass operator.
\end{abstract}

\tableofcontents

%%%%%%%%%%%%%%%%%%%%%%%%%%%%%%%%%%%%%%%%%%%%%%%%%%%%%%%%%%%%%%%%%%%%%%
%%%%%%%%%%%%%%%%%%%%%%%%%%%%%%%%%%%%%%%%%%%%%%%%%%%%%%%%%%%%%%%%%%%%%%
\section{Introduction}
%%%%%%%%%%%%%%%%%%%%%%%%%%%%%%%%%%%%%%%%%%%%%%%%%%%%%%%%%%%%%%%%%%%%%%
%%%%%%%%%%%%%%%%%%%%%%%%%%%%%%%%%%%%%%%%%%%%%%%%%%%%%%%%%%%%%%%%%%%%%%

Certainly, the available computer power has advanced impressively
over the last years. Nevertheless, for obtaining high precision simulation
results in lattice QCD, as our target application in this paper, it remains
essential to improve --on the one hand-- the algorithms employed for lattice
simulations and --on the other hand-- to find better formulations of lattice
fermions.  Two very promising candidates for improved versions of lattice
fermions are chirally {\em invariant} \cite{Luscher:1998pq} overlap fermions
\cite{Neuberger:1997fp, Neuberger:1998wv} and chirally {\em improved} Wilson
twisted mass (TM) fermions \cite{Frezzotti:2003ni} at maximal twist.  Both
have the potential to overcome some basic difficulties of lattice QCD, most
notably they make simulations at values of the pseudo scalar mass
close to the experimentally observed pion mass of $140$MeV
possible. For a comparison of physical results obtained with the two 
mentioned operators in the quenched approximation see
Ref.~\cite{Bietenholz:2004wv}.

The reason for these difficulties is that one has to solve a huge set of
linear equations over and over again. Although, due to the only nearest
neighbour interaction of the underlying Wilson-Dirac operator, sparse matrix
methods can be employed, the computational cost can get extremely large, see,
e.g. the discussions in Refs.~\cite{Ukawa:2002pc,Jansen:2003nt,Urbach:2005ji}.

The focus of this work is to compare different iterative linear
solvers\footnote{What is needed for lattice calculations are certain rows or
  columns of the {\em inverse} of the fermion matrix employed which are
  obtained from the solution of a set of linear equations. By abuse of
  language, we will therefore sometimes speak about ``inversion algorithms'',
  ``inverse operator'' etc.~while the mathematical problem is always the
  solution of a large set of linear equations using iterative sparse matrix
  methods.}  for sparse matrices as needed for computing the quark propagator
for valence quarks or for the computation of the fermionic ''force'' in
dynamical simulations.  It has to be remarked that the exact behaviour of
sparse matrix methods is highly problem specific and can depend strongly on
the underlying matrix involved. It is hence crucial to compare the optimal
method for a given kind of lattice fermion. In our case, we will consider
overlap fermions and Wilson twisted mass fermions at maximal twist.  We will
explore a number of sparse matrix methods for the solution of the linear
system defined by the corresponding lattice Dirac operator. Although we have
tried to be rather comprehensive, it is clear that such a work cannot be
exhaustive. The set of possible linear solvers is too large to be able to
cover all of them, see e.g. \cite{saad:2003a} and different solvers may be
better for different situations. For example, if we are only interested in
computing fermion propagators, the question is, whether we want to have a
multiple mass solver \cite{Frommer:1995ik}.  Or, with respect of dynamical
simulations, we need the square of the lattice Dirac operator and not the
operator itself which can lead to very different behaviour of the algorithm
employed.  In addition, each of the basic algorithms can be combined with
certain improvement techniques which again influence the algorithm behaviour
substantially.

In principle, it is also desirable to study the performance
behaviour of the algorithms as a function of the pseudo scalar mass, the
lattice volume and the lattice spacing. Again, in this work, due to the very
costly simulations such a study would require, we have to restrict ourselves
to an only limited set of parameters. In particular, we will consider two
physical volumes and four values of the pseudo scalar mass (matched between
both formulations of lattice QCD).  Finally, we will only take one value of
the lattice spacing for our study.

As an outcome of this work, we will find that the computational cost of
particular algorithms and variants thereof can vary substantially for
different situations.  This gives rise to the conclusion that it can be very
profitable to test the --at least most promising-- algorithms for the
particular problem one is interested in. It is one of our main conclusions
that easily a factor of two or larger can be gained when the algorithm is
adopted to the particular problem under consideration.

Parts of the results presented in this paper were already published in
Ref.~\cite{Chiarappa:2004ry} and for related work concerning the
overlap operator see
Refs.~\cite{Arnold:2003sx,Cundy:2004pz,Krieg:2004xg}.

The outline of the paper is as follows. In section \ref{sec:Lattice Dirac
  operators} we introduce the Dirac operators that are considered in this
study.  Section \ref{sec:Iterative linear solver algorithms} discusses the
iterative linear solver algorithms and special variants thereof, like multiple
mass solvers and, for the overlap operator, adaptive precision solvers and
solvers in a given chiral sector. In section \ref{sec:Preconditioning
  techniques} we present various preconditioning techniques like even/odd
preconditioning for the TM operator and low mode preconditioning for the
overlap operator. In section \ref{sec:Results} we present and discuss our
results and in section \ref{sec:Conclusions and Outlook} we finish with
conclusions and an outlook.  Appendix \ref{app:ev} deals with the computation
of eigenvalues and eigenvectors which is important for an efficient
implementation of the overlap operator, for its low mode preconditioning and
for the computation of its index. Appendix \ref{app:MM for TM} finally
reformulates the TM operator so that multiple mass solvers become applicable.

%%%%%%%%%%%%%%%%%%%%%%%%%%%%%%%%%%%%%%%%%%%%%%%%%%%%%%%%%%%%%%%%%%%%%%
%%%%%%%%%%%%%%%%%%%%%%%%%%%%%%%%%%%%%%%%%%%%%%%%%%%%%%%%%%%%%%%%%%%%%%
\section{Lattice Dirac operators} 
\label{sec:Lattice Dirac operators}
%%%%%%%%%%%%%%%%%%%%%%%%%%%%%%%%%%%%%%%%%%%%%%%%%%%%%%%%%%%%%%%%%%%%%%
%%%%%%%%%%%%%%%%%%%%%%%%%%%%%%%%%%%%%%%%%%%%%%%%%%%%%%%%%%%%%%%%%%%%%%
We consider QCD on a four-dimensional hyper-cubic lattice in Euclidean
space-time. The fermionic fields $\psi$ live on the sites $x$ of the lattice
while the SU(3) gauge fields of the theory are represented by group-valued
link variables $U_\mu(x), \mu=1,\ldots,4$. The gauge covariant backward and
forward difference operators are given by
\begin{equation}
  \begin{split}
    (\nabla_\mu \psi)(x) &= U_\mu(x) \psi(x+\hat\mu) - \psi(x), \\
    (\nabla^*_\mu \psi)(x) &= \psi(x) - U^\dagger_\mu(x-\hat\mu) \psi(x-\hat\mu),
  \end{split}
\end{equation}
and the standard Wilson-Dirac operator with bare quark mass $m_0$ can be
written as
\begin{equation}
  \label{eq:Dw}
  D_\text{W}(m_0) = \sum_{\mu=1}^4 \frac{1}{2} \{\gamma_\mu (\nabla_\mu
  + \nabla^*_\mu) - \nabla^*_\mu \nabla_\mu \} + m_0 . 
\end{equation}

The twisted mass lattice Dirac operator for a $SU_f(2)$ flavour doublet of
mass degenerate quarks has the form \cite{Frezzotti:1999vv,Frezzotti:2000nk}
\begin{equation}
  \label{eq:Dtm}
  D_{\text{tm}}(\mu_\mathrm{tm}) = D_\text{W}(m_0) + 
  i \mu_\mathrm{tm} \gamma_5  \tau_3\, ,
\end{equation}
where $D_\text{W}$ is the Wilson-Dirac operator with bare quark mass $m_0$ as
defined above, $\mu_\mathrm{tm}$ the twisted quark mass and $\tau_3$ the third
Pauli matrix acting in flavour space. Since it was shown in
Ref.~\cite{Frezzotti:2003ni} (for a test in practise see
Refs.~\cite{Jansen:2003ir,Jansen:2005kk}) that physical observables are
automatically $O(a)$ improved if $m_0$ is tuned to its critical value, we are
only interested in this special case.

The second operator we consider, the massive overlap operator, is
defined as \cite{Neuberger:1997fp,Neuberger:1998wv}
\begin{equation}
  \label{eq:Dov}
  D(\mu_\mathrm{ov}) = 
  \Big(1 - \frac{\mu_\mathrm{ov}}{2 M}\Big) D + \mu_\mathrm{ov}\, ,
\end{equation}
where 
\begin{equation}
  \label{eq:Dovmassless}
  D = M \Big(1 + 
  \gamma_5 \sign\left[Q(-M)\right]\Big)\,
\end{equation}
is the massless overlap operator, $Q(-M) = \gamma_5 D_\text{W}(-M)$ with $M$
chosen to be $M=1.6$ in this work and $\mu_\mathrm{ov}$ again the bare quark
mass. The matrix $\sign$-function in Eq.(\ref{eq:Dovmassless}) is calculated
by some approximation that covers the whole spectrum of $Q(-M)$. To make this
feasible we determine $K$ eigenmodes of $Q(-M)$ closest to the origin,
project them out from the $\sign$-function and calculate their contribution
analytically while the rest of the spectrum is covered by an approximation
employing Chebysheff polynomials. Denoting by $\psi_k$ the eigenvectors of $Q$
with corresponding eigenvalue $\lambda_k$, i.e.~$Q \psi_k = \lambda_k \psi_k$,
we have
\begin{equation}
  \label{eq:sign}
  \begin{split}
      \sign \left[Q(-M)\right] =& \sum_{k=1}^K \sign(\lambda_k) P_k +\\
      & \left(1-\sum_{k=1}^K P_k\right) Q \cdot T_N\left[Q^2\right]
      \left(1-\sum_{k=1}^K P_k\right)\, , \\ 
  \end{split}
\end{equation}
where $P_k = \psi_k \psi_k^\dagger$ are projectors onto the eigenmode
subspaces and $T_N\left[Q^2\right]$ denotes the $N$-th order Chebysheff
polynomial approximation to $1/\sqrt(Q^2)$ on the orthogonal subspace.  The
calculation of the eigenmodes is discussed in appendix~\ref{app:ev}.

%%%%%%%%%%%%%%%%%%%%%%%%%%%%%%%%%%%%%%%%%%%%%%%%%%%%%%%%%%%%%%%%%%%%%%
%%%%%%%%%%%%%%%%%%%%%%%%%%%%%%%%%%%%%%%%%%%%%%%%%%%%%%%%%%%%%%%%%%%%%%
\section{Iterative linear solver algorithms}
\label{sec:Iterative linear solver algorithms}
%%%%%%%%%%%%%%%%%%%%%%%%%%%%%%%%%%%%%%%%%%%%%%%%%%%%%%%%%%%%%%%%%%%%%%
%%%%%%%%%%%%%%%%%%%%%%%%%%%%%%%%%%%%%%%%%%%%%%%%%%%%%%%%%%%%%%%%%%%%%%
Let us now turn to the iterative linear solver algorithms that we consider in
our investigation. Table~\ref{tab:alglist} lists the various algorithms and
marks with 'x' which of them are used with the overlap and the TM operator,
respectively. For the convenience of the reader we also compile in table
\ref{tab:alglist} for each algorithm the number of operator applications,
i.e.~matrix-vector (MV) multiplications, together with the corresponding
number of scalar products (SP) and linear algebra instructions $Z = \alpha X +
Y$ (ZAXPY) per iteration. Moreover, in the last column we also note which of
the algorithms possess the capability of using multiple masses (MM).
\begin{table}[thb]
\centering
    \begin{tabular*}{1.\textwidth}{@{\extracolsep{\fill}}lcccccc}
      \hline
      $\Bigl.\Bigr.$Algorithm \hspace*{6mm}& Overlap & TM & MV & SP & ZAXPY & MM \\
      \hline\hline
      CGNE \cite{saad:2003a}    & x & x & 2 & 2       & 3 & yes \\
      CGS \cite{saad:2003a}       & x & x & 2 & 2       & 7 & yes    \\
      BiCGstab \cite{saad:2003a}  & x & x & 2 & 4       & 6 & yes \\
      GMRES(m) \cite{saad:2003a}  & x & x & 1 & $m/2+1$ & $m/2+1$ & no\\
      MR \cite{saad:2003a}        & x & x & 1 & 2       & 2 & yes \\
      CGNE$_\chi$         & x &   & 1 & 2       & 3 & yes \\
      SUMR \cite{Jagels:1994a}    & x &   & 1 & 6       & 1 & yes  \\
      \hline
    \end{tabular*}
    \caption{Linear solver algorithms for the overlap and twisted mass (TM)
      operator. Also given are the number of matrix vector (MV) multiplications,
      scalar products (SP) and $z=\alpha x+y$ (ZAXPY) linear algebra operations
      per iteration. We also indicate, whether the algorithm can be used to solve 
      for multiple masses (MM).}
  \label{tab:alglist}
\end{table}

With the exception of the MR algorithm all algorithms are
Krylov subspace methods, i.e.~they construct the solution of the linear system
$A \psi = \eta$ as a linear combination of vectors in the Krylov subspace
\[
\mathcal{K}_i = \text{span}(v,Av,\ldots, A^{i-1}v)\, ,
\]
where $v=r_0= \eta-A\psi_0$ is the initial residual. In contrast the MR
algorithm is a one-dimensional projection process \cite{saad:2003a}, i.e.~each
iteration step is completely independent of the previous one.  For a detailed
description and discussion of the basic algorithms we refer to
\cite{saad:2003a}, whereas we will discuss some special versions in the
following subsections. Note that we adopted the names of the algorithms from
Ref.~\cite{Saad:1993a} where possible.

The SUMR algorithm was introduced in Ref.~\cite{Jagels:1994a} and first used
for lattice QCD in Ref.~\cite{Arnold:2003sx}. It makes use of the unitarity
property of the massless overlap operator and was shown to perform rather well
when compared to other standard iterative solvers \cite{Arnold:2003sx}.

In the case of TM fermions it is sometimes useful to consider the linear
system $\gamma_5 A \psi = \gamma_5\eta$ instead of $A \psi = \eta$.  The
reason for the importance of this change will be discussed later. We will add
a $\gamma_5$ to the solver name in case the changed system is solved, like for
instance CGS$\gamma_5$.

%%%%%%%%%%%%%%%%%%%%%%%%%%%%%%%%%%%%%%%%%%%%%%%%%%%%%%%%%%%%%%%%%%%%%%
\subsection{Multiple mass solvers}
%%%%%%%%%%%%%%%%%%%%%%%%%%%%%%%%%%%%%%%%%%%%%%%%%%%%%%%%%%%%%%%%%%%%%%
In propagator calculations for QCD applications it is often necessary to
compute solutions of the system
\begin{equation}
  \label{eq:shiftedsystem}
  (A+\sigma)\psi = \eta
\end{equation}
for several values of a scalar shift $\sigma$ - usually the mass. It has been
realised some time ago
\cite{Frommer:1995ik,Glassner:1996gz,Jegerlehner:1996pm,Jegerlehner:2003qp}
that the solutions of 
the shifted systems can be obtained at largely the cost of only solving the
system with the smallest (positive) shift. For the Krylov space solvers this
is achieved by realising that the Krylov spaces of the shifted systems are
essentially the same. In table \ref{tab:alglist} we note in the last column
which of the algorithms can be implemented with multiple masses.  Multiple
mass (MM) versions for BiCGstab and CG can be found in
\cite{Jegerlehner:1996pm}.  In principle there exists also a MM version for
the GMRES algorithm, but since in practise the GMRES has to be restarted after
$m$ iteration steps it does not carry over to the case of GMRES($m$). For the
SUMR algorithm we note that the MM version is trivial, since the shift of the
unitary matrix enters in the algorithm not via the iterated vectors but
instead only through scalar coefficients directly into the solution
vector.

Finally we wish to emphasise that also the CGNE algorithm is capable of using
multiple masses in special situations. This remark is non-trivial since in
general $(A^\dagger + \sigma)(A+\sigma)$ appearing in the normal equation is
not of the form $A'^\dagger A' + \sigma'$. However, it turns out that for the
overlap operator and the twisted mass operator a MM is possible. For the
overlap operator one can make use of the Ginsparg-Wilson relation in order to
bring the shifted normal equation operator into the desired form
\cite{Edwards:1998wx}.  For the Wilson twisted mass fermion operator we
provide in Appendix \ref{app:MM for TM} the details of the MM implementation
for CGNE.

%%%%%%%%%%%%%%%%%%%%%%%%%%%%%%%%%%%%%%%%%%%%%%%%%%%%%%%%%%%%%%%%%%%%%%
\subsection{Chiral CGNE for the overlap operator}
%%%%%%%%%%%%%%%%%%%%%%%%%%%%%%%%%%%%%%%%%%%%%%%%%%%%%%%%%%%%%%%%%%%%%%
Due to the fact that the overlap operator obeys the Ginsparg-Wilson relation
it is easy to show that $D^\dagger D$ commutes with $\gamma_5$. As a
consequence the solution to the normal equation $D^\dagger D \psi = \eta$ can
be found in a given chiral sector as long as the original source vector $\eta$
is chiral. (This is for example the case if one works with point sources in a
chiral basis.)

When applying the CGNE algorithm to the overlap operator one can then make use
of this fact by noting the relation
\begin{equation}
  \label{eq:PDP operator}
  P_\pm D(\mu_{\text{ov}})^\dagger D(\mu_{\text{ov}}) P_\pm = 2 M
  P_\pm D(\mu_{\text{ov}}^2/(2M)) P_\pm\, ,
\end{equation} 
where $P_\pm=1/2(1\pm\gamma_5)$ are the chiral projectors. Thus in each
iteration the operator is only applied once instead of twice, but with a
modified mass parameter. This immediately saves a factor of two in the number
of matrix-vector (MV) applications with respect to the general case. In table
\ref{tab:alglist} and in the following we denote this algorithm by
CGNE$_\chi$.

%%%%%%%%%%%%%%%%%%%%%%%%%%%%%%%%%%%%%%%%%%%%%%%%%%%%%%%%%%%%%%%%%%%%%%
\subsection{Adaptive precision solvers for the overlap operator}
\label{subsec:adaptive precision}
%%%%%%%%%%%%%%%%%%%%%%%%%%%%%%%%%%%%%%%%%%%%%%%%%%%%%%%%%%%%%%%%%%%%%%
It is well known that the computational bottleneck for the solvers employing
the overlap operator is the computation of the approximation of the
sign-function $\text{sign}(Q)$. Since each application of the overlap operator
during the iterative solver process requires yet another iterative procedure
to approximate $\sign(Q)$, we are led to
a two-level nested iterative procedure where the cost for the calculation of
the sign-function enters multiplicatively in the total cost. So any optimised
algorithm will not only aim at minimising the number of outer iterations,
i.e.~the number of overlap operator applications, but it will also try to
reduce the number of inner iterations, i.e.~the order of the -- in our case
polynomial -- approximation.

While the problem of minimising the number of outer iterations depends on a
delicate interplay between the algorithm and the operator under consideration
and comprises one of the main foci of the present investigation, the problem
to reduce the number of inner iterations can be achieved rather directly in
two different ways. Firstly, as discussed in section \ref{sec:Lattice Dirac
  operators}, we project out the lowest 20 or 40 eigenvectors of the
Wilson-Dirac operator depending on the extent of the lattice
(cf.~section 
\ref{subsec:Setup}). In this way we achieve that our approximations use
(Chebysheff) polynomials typically of the order $O(200-300)$ for the
simulation parameters we have employed for this study.

Secondly, it is then also clear that one can speed up the calculations by
large factors if it is possible to reduce the accuracy of the approximation.
In realising this, the basic idea is to {\em adapt} the degree of the
polynomial during the solver iteration to achieve only that precision as
actually needed in the present iteration step.  We have implemented the
adaptive precision for a selection of the algorithms that seemed most
promising in our first tests and in the following we denote these algorithms
by the subscript $_\text{ap}$ for adaptive precision.  Usually not more than
two lines of additional code are required to implement the adaptive precision
versions of the algorithms. Obviously the details of how exactly one needs to
adapt the precision of the polynomial depends on the details of the algorithm
itself and might also influence the possibility to do multi mass inversions.

We use two generic approaches which we illustrate in the following by means of
the adaptive precision versions of the MR and the CGNE algorithms,
respectively. In the case of the MR we follow a strategy that is similar to
restarting: through the complete course of the iterative procedure we use a
low order polynomial approximation of a degree $O(10)$ for the $\sign$
function. Only every $m$ iteration steps we correct for the errors by
computing the true residuum to full precision, which corresponds essentially
to a restart of the algorithm. We denote this algorithm with
MR$_\mathrm{ap}(m)$.  We remark that with this approach the MM capability of
the MR algorithm is lost.

The MR$_\mathrm{ap}(m)$ is outlined with pseudo-code in algorithm \ref{mrap},
where we denote the low order approximation of the overlap operator with
$A_\mathrm{ap}$ while the full precision operator is denoted with $A$.

\begin{algorithm}
  \caption{MR$_\mathrm{ap}(A, A_\mathrm{ap}, b, x, m, \epsilon)$ algorithm}
  \label{mrap}
  \begin{algorithmic}[1]
    \STATE $i=0$
    \STATE $p = Ax$
    \STATE $r = b - p$
    \REPEAT
    \STATE $i=i+1$
    \STATE \COMMENT{Use $A_\mathrm{ap}$ with fixed low order polynomial}
    \STATE $\tilde r = A_\mathrm{ap} r$ 
    \STATE $\alpha = (\tilde r,r)/(\tilde r, \tilde r)$
    \STATE $x = x + \alpha r$
    \IF{$i\mod m = 0$}
    \STATE \COMMENT{Correct with full $A$}
    \STATE $p = Ax$ 
    \ELSE
    \STATE $p = p + \alpha \tilde r$
    \ENDIF
    \STATE $r = b - p$
    \UNTIL{$\|r\| < \epsilon$}
  \end{algorithmic}
\end{algorithm}

The same approach as used for the MR$_\mathrm{ap}(m)$ algorithm can easily be
carried forward to the GMRES$_\mathrm{ap}(m)$ algorithm. Since the GMRES$(m)$
is restarted every $m$ iterations, we use only every $m$-th iteration the full
approximation to the $\sign$ function while all other applications of the
overlap operator are performed with an approximation of degree $O(10)$.

In case of the CGNE$_\mathrm{ap}$ our strategy is different: here we simply
calculate contributions to the $\sign$-function approximation up to the point
where they are smaller than $\epsilon_\mathrm{ap} = 10^{-2}\epsilon$, where
$\epsilon$ is the desired final residual, i.e.~we neglect all corrections that
are much smaller than the final residual. This requires the full polynomial
only at the beginning of the CG-search while towards the end of the search we
use polynomials with a degree $O(10)$. In order to implement this idea we use
a forward recursion scheme for the application of the Chebysheff polynomial as
detailed in algorithm \ref{chap}.

\begin{algorithm}
  \caption{Compute $r=\sum_{j=0}^{n-1} c_j T_j(Q^2)\, v$ to precision $\epsilon_\mathrm{ap}$}
  \label{chap}
  \begin{algorithmic}[1]
    \REQUIRE vector $v$ and Chebysheff coefficients $c_j$
    \vspace{.2cm}
    \STATE $d_0 = T_0(Q^2)\, v = v$
    \STATE $d_1 =  T_1(Q^2)\, v = 2\, Q^2\, v - v$
    \STATE $r = c_1 d_1 + 1/2 c_0 d_0$
    \FOR{j=2,...,n-1}
    \STATE $d_j = T_j(Q^2)\, v = 2\, Q^2\, d_{j-1} - d_{j-2} $
    \STATE $r = r + c_j d_j$
    \IF{$\|d_j\| < \epsilon_\mathrm{ap}$}
    \STATE return $r$
    \ENDIF
    \ENDFOR
    \STATE return $r$
  \end{algorithmic}
\end{algorithm}

It is important to note here that with this approach for the CGNE$_\text{ap}$
the MM capability is preserved (in contrast to an approach proposed in
Ref.~\cite{Giusti:2002sm} similar to the MR$_\text{ap}(m)$ approach described
above where the MM capability is lost). The strategy for the SUMR$_\text{ap}$
is analogous to the one for the CGNE$_\text{ap}$, where again the MM
capability is preserved.

%%%%%%%%%%%%%%%%%%%%%%%%%%%%%%%%%%%%%%%%%%%%%%%%%%%%%%%%%%%%%%%%%%%%%%
%%%%%%%%%%%%%%%%%%%%%%%%%%%%%%%%%%%%%%%%%%%%%%%%%%%%%%%%%%%%%%%%%%%%%%
\section{Preconditioning techniques}
\label{sec:Preconditioning techniques}
%%%%%%%%%%%%%%%%%%%%%%%%%%%%%%%%%%%%%%%%%%%%%%%%%%%%%%%%%%%%%%%%%%%%%%
%%%%%%%%%%%%%%%%%%%%%%%%%%%%%%%%%%%%%%%%%%%%%%%%%%%%%%%%%%%%%%%%%%%%%%
\subsection{Even/odd preconditioning for the TM operator}
\label{subsec:eo}
%%%%%%%%%%%%%%%%%%%%%%%%%%%%%%%%%%%%%%%%%%%%%%%%%%%%%%%%%%%%%%%%%%%%%%
Even/odd preconditioning for the Wilson TM operator has already been described
in \cite{Farchioni:2004us} and we review it here for completeness only. Let us
start with the hermitian two flavour Wilson TM operator\footnote{In this
  section we suppress the subscript ${}_\mathrm{tm}$ for notational
  convenience and simply write $D$ for $D_\mathrm{tm}$ and $\mu$ for
  $\mu_\mathrm{tm}$.} in the hopping parameter representation
($\kappa=(2m_0+8)^{-1}$)
\begin{equation}
  \label{eq:eo1}
  Q\equiv \gamma_5 D = \begin{pmatrix}
    \Qp & \\\
    & \Qm \\
  \end{pmatrix}\, ,
\end{equation}
where the sub-matrices $\Qpm$ can be factorised with
$\tilde\mu=2\kappa\mu$ as follows:
\begin{equation}
  \label{eq:eo2}
  \begin{split}
    Q^\pm &= \gamma_5\begin{pmatrix}
      1\pm i\tilde\mu\gamma_5 & D_{eo} \\
      D_{oe}    & 1\pm i\tilde\mu\gamma_5 \\
    \end{pmatrix} =
    \gamma_5\begin{pmatrix}
      D_{ee}^\pm & D_{eo} \\
      D_{oe}    & D_{oo}^\pm \\
    \end{pmatrix} \\
    & =
    \begin{pmatrix}
      \gamma_5D_{ee}^\pm & 0 \\
      \gamma_5D_{oe}  & 1 \\
    \end{pmatrix}
    \begin{pmatrix}
      1       & (D_{ee}^\pm)^{-1}D_{eo}\\
      0       & \gamma_5(D_{oo}^\pm-D_{oe}(D_{ee}^\pm)^{-1}D_{eo})\\
    \end{pmatrix}\, .
\end{split}
\end{equation}
Note that $(D_{ee}^\pm)$ is trivial to invert:
\begin{equation}
  \label{eq:eo3}
  (1\pm i\tilde\mu\gamma_5)^{-1} = \frac{1\mp i\tilde\mu\gamma_5}{1+\tilde\mu^2}.
\end{equation}
Due to the factorisation (\ref{eq:eo2}) the full fermion matrix can be
inverted by inverting the two matrices appearing in the factorisation
\[
\begin{pmatrix}
  D_{ee}^\pm & D_{eo} \\
  D_{oe}    & D_{oo}^\pm \\
\end{pmatrix}^{-1}
=
\begin{pmatrix}
  1       & (D_{ee}^\pm)^{-1}D_{eo}\\
  0       & (D_{oo}^\pm-D_{oe}(D_{ee}^\pm)^{-1}D_{eo})\\
\end{pmatrix}^{-1}
\begin{pmatrix}
  D_{ee}^\pm & 0 \\
  D_{oe}   & 1 \\
\end{pmatrix}^{-1}\, .
\]
and the two factors can be simplified as follows:
\[
\begin{pmatrix}
  D_{ee}^\pm & 0 \\
  D_{oe}   & 1 \\
\end{pmatrix}^{-1}
=
\begin{pmatrix}
      (D_{ee}^\pm)^{-1} & 0 \\
      -D_{oe} (D_{ee}^{\pm})^{-1}  & 1 \\
    \end{pmatrix}
\]
and 
\[
\begin{split}
  &\begin{pmatrix}
    1       & (D_{ee}^\pm)^{-1}D_{eo}\\
    0       & (D_{oo}^\pm-D_{oe}(D_{ee}^\pm)^{-1}D_{eo})\\
  \end{pmatrix}^{-1}
  \\=&
  \begin{pmatrix}
    1       & -(D_{ee}^\pm)^{-1}D_{eo}(D_{oo}^\pm-D_{oe}(D_{ee}^\pm)^{-1}D_{eo})^{-1}  \\
    0       & (D_{oo}^\pm-D_{oe}(D_{ee}^\pm)^{-1}D_{eo})^{-1}\\
  \end{pmatrix}\, .
\end{split}
\]
The complete inversion is now performed in two separate steps: First
we compute for a given source field $\phi=(\phi_e,\phi_o)$ an intermediate 
result $\varphi=(\varphi_e,\varphi_o)$ by:
\[
\begin{pmatrix}
  \varphi_e \\ \varphi_o\\
\end{pmatrix}
=
\begin{pmatrix}
  D_{ee}^\pm & 0 \\
  D_{oe}   & 1 \\
\end{pmatrix}^{-1}
\begin{pmatrix}
  \phi_e \\ \phi_o \\
\end{pmatrix}
=
\begin{pmatrix}
  (D_{ee}^\pm)^{-1} \phi_e \\ 
  -D_{oe}( D_{ee}^\pm)^{-1} \phi_e + \phi_o \\
\end{pmatrix}\, .
\]
This step requires only the application of $D_{oe}$ and $(D_{ee}^\pm)^{-1}$,
the latter of which is given by Eq.(\ref{eq:eo3}).  The final solution
$\psi=(\psi_e,\psi_o)$ can then be computed with
\[
\begin{pmatrix}
  \psi_e \\ \psi_o \\
\end{pmatrix}
=
\begin{pmatrix}
  1       & (D_{ee}^\pm)^{-1}D_{eo}\\
  0       & (D_{oo}^\pm-D_{oe}(D_{ee}^\pm)^{-1}D_{eo})\\
\end{pmatrix}^{-1}
\begin{pmatrix}
  \varphi_e \\ \varphi_o \\
\end{pmatrix}
=
\begin{pmatrix}
  \varphi_e - (D_{ee}^\pm)^{-1}D_{eo}\psi_o \\ \psi_o \\
\end{pmatrix}\, ,
\]
where we defined
\begin{equation}
  \label{eq:eo4}
  \psi_o = (D_{oo}^\pm-D_{oe}(D_{ee}^\pm)^{-1}D_{eo})^{-1}
  \varphi_o\,=\hat D^{-1} \varphi_o\, . 
\end{equation}
Therefore the only inversion that has to be performed numerically is the one
to generate $\psi_o$ from $\varphi_o$ and this inversion involves only $\hat
D$ that is better conditioned than the original fermion operator.

A similar approach is to invert in Eq.(\ref{eq:eo4}) instead of $\hat D$ the
following operator:
\[
\hat D_s = 1-(D_{oo}^\pm)^{-1}D_{oe}(D_{ee}^\pm)^{-1}D_{eo}\, ,
\]
on the source $(D_{oo}^\pm)^{-1} \varphi_o$. As noticed already in
Ref.~\cite{Jansen:1997yt} for the case of non-perturbatively improved Wilson
fermions this more symmetrical treatment results in a slightly better
condition number leading to $20\%$ less iterations in the solvers.

%%%%%%%%%%%%%%%%%%%%%%%%%%%%%%%%%%%%%%%%%%%%%%%%%%%%%%%%%%%%%%%%%%%%%%
\subsection{Low mode preconditioning for the overlap operator}
\label{subsec:lmp overlap}
%%%%%%%%%%%%%%%%%%%%%%%%%%%%%%%%%%%%%%%%%%%%%%%%%%%%%%%%%%%%%%%%%%%%%%
Low mode preconditioning (LMP) for the overlap operator has already been
described in Ref.~\cite{Giusti:2002sm} using the CG algorithm on the normal
equations. In case of the CG the operator $D^\dagger D$ to be inverted
is hermitian, and hence normal, and the low mode preconditioning is as
described in Ref.~\cite{Giusti:2002sm}.

The application of this technique to algorithms like GMRES or MR (which
involve $D$ instead of $D^\dagger D$) is not completely straightforward.
Although the overlap operator itself is formally normal, in practise it is not
due to the errors introduced by the finite approximation of the
$\sign$-function\footnote{Note that for the CGNE algorithm used in
  \cite{Giusti:2002sm} the non-normality of the approximate overlap operator
  is circumvented by construction since $D^\dagger D$ is hermitian for any
  approximation of the $\sign$-function.}.  As a consequence one has to
distinguish between left and right eigenvectors of $D$ leading to some
additional complications which we are now going to discuss.
 
Consider the linear equation $A \psi = \eta$. The vector space on which the
linear operator $A$ acts can be split into two (bi-)orthogonal pieces using
the (bi-)orthogonal projectors
\begin{equation} 
P = \sum_k r_k l_k^\dagger, \quad P_\perp = 1 - P.
\end{equation}
Here we assume that the $r_k's$ and $l_k's$ are approximate right and left
eigenvectors (Ritz vectors), respectively, of the operator $A$ which form a
bi-orthogonal basis, i.e.~$l_i^\dagger r_j = \delta_{ij}$. One can write
\begin{eqnarray}\label{eq:right gradient}
A r_k &=& \alpha_k r_k + g_k^{(r)}\, , \\
\label{eq:left gradient}
A^\dagger l_k &=& \bar \alpha_k l_k + g_k^{(l)}\, ,
\end{eqnarray}
where $l_i^\dagger g_k^{(r)} = r_i^\dagger g_k^{(l)} = 0$. Indeed, one finds
\begin{equation}
P A r_k = \alpha_k r_k 
\end{equation}
and
\begin{equation}
P_\perp A r_k = g_k^{(r)}.
\end{equation}
The operator $A$ then takes the following block form
\begin{equation} 
A = \left( 
\begin{array}{cc}
P A P & P A P_\perp \\
P_\perp A P & P_\perp A P_\perp
\end{array}
\right)
\end{equation}
and the linear equation reads
\begin{equation} 
\left( 
\begin{array}{cc}
P A P & P A P_\perp \\
P_\perp A P & P_\perp A P_\perp
\end{array}
\right)
\left( 
\begin{array}{c}
P \psi \\
P_\perp \psi
\end{array}
\right)
=
\left( 
\begin{array}{c}
P \eta \\
P_\perp \eta
\end{array}
\right).
\end{equation}
To solve this equation we can perform a LU decomposition of $A$
\begin{equation} 
A =
\left( 
\begin{array}{cc}
1 & 0 \\
P_\perp A P (P A P)^{-1} & 1
\end{array}
\right)
\left( 
\begin{array}{cc}
P A P & P A P_\perp \\
 0    & S 
\end{array}
\right) \equiv L \cdot U\, ,
\end{equation}
where $S = P_\perp A P_\perp - P_\perp A P (P A P)^{-1} P A P_\perp$ is the
Schur complement of $A$. The lower triangular matrix $L$ can be
inverted and applied to the right hand side,
\begin{equation} 
L^{-1} \eta =
\left( 
\begin{array}{c}
P \eta  \\
- P_\perp A P (P A P)^{-1} P \eta  + P_\perp \eta
\end{array}
\right),
\end{equation}
and the linear system reduces to solving $U (P \psi, P_\perp \psi)^T =
L^{-1}\eta$. Written out explicitly we obtain the second component $P_\perp
\psi$ from solving the equation
\begin{equation}\label{eq:P_perp psi}
P_\perp (A  - A P (P A P)^{-1} P A ) P_\perp \psi 
= P_\perp \eta - P_\perp A P (P A P)^{-1} P \eta
\end{equation}
and the first component $P \psi$ from the solution of 
\begin{equation}
P A P \cdot P \psi = P \eta - P A P_\perp \psi.
\end{equation}

In detail the whole procedure to solve $A \psi = \eta$ using low mode
preconditioning involves the following steps:
\begin{enumerate}
\item prepare (precondition) the source according to the r.h.s.~of
  Eq.(\ref{eq:P_perp psi}), i.e.
  \begin{equation}
    \label{eq:lmp0}
    \eta' = P_\perp \left( 1 - \sum_i g_i^{(r)} \frac{1}{\alpha_i}
      l_i^\dagger \right) \eta\, ,
  \end{equation}
  where we have used $P_\perp r_i = 0$.
\item solve the low mode preconditioned system $A_\text{lmp} P_\perp \psi =
  \eta'$ for $P_\perp \psi$ where $A_\text{lmp}$ is the preconditioned
  operator acting in the subspace orthogonal to the low modes, i.e.~the
  operator on the l.h.s.~of Eq.(\ref{eq:P_perp psi}). To be specific the
  application of the preconditioned operator is given by
  \begin{multline}
    \label{eq:A_lmp}
    P_\perp \left[ A - A P (P A P)^{-1} P A ) \right] P_\perp \psi = \\
    P_\perp \left[ A - \sum_{i,j,k} \left( \alpha_i r_i + g_i^{(r)} \right)  l_i^\dagger r_j
      \frac{1}{\alpha_j} l_j^\dagger r_k \left( l_k^\dagger \alpha_k  + g_k^{(l)}{}^\dagger
      \right) \right] P_\perp \psi \\
     =P_\perp \left[ A - \sum_i g_i^{(r)} \frac{1}{\alpha_i} g_i^{(l)}{}^\dagger
    \right] P_\perp \psi \, ,
  \end{multline}
  where we have used $P_\perp r_i = l_k^\dagger P_\perp = 0$.
\item add in the part of the solution from the subspace spanned by the low
  modes, i.e.~$P \psi$. This part is essentially the contribution from the low
  modes and it is explicitly given by
  \begin{equation}
    \label{eq:P psi}
    P \psi = \sum_i r_i \frac{1}{\alpha_i} l_i^\dagger (\eta - A P_\perp \psi).
  \end{equation}
\end{enumerate}

Let us mention for completeness that there are further related preconditioning
techniques available which do not involve the analytic correction step in
Eq.(\ref{eq:P psi}). The Ritz vectors can be used directly in any right or
left preconditioned version of a given solver like for instance in the FGMRES
algorithm \cite{Saad:1993a}. Moreover, the computation of the Ritz pairs and
the iterative solution can be combined in so called iterative solvers with
deflated eigenvalues, see for instance the GMRES versions discussed in
Refs.\cite{saad:2003a,Burrage:1998a,Morgan:2002a}.

%%%%%%%%%%%%%%%%%%%%%%%%%%%%%%%%%%%%%%%%%%%%%%%%%%%%%%%%%%%%%%%%%%%%%%
%%%%%%%%%%%%%%%%%%%%%%%%%%%%%%%%%%%%%%%%%%%%%%%%%%%%%%%%%%%%%%%%%%%%%%
\section{Results}
\label{sec:Results} 
%%%%%%%%%%%%%%%%%%%%%%%%%%%%%%%%%%%%%%%%%%%%%%%%%%%%%%%%%%%%%%%%%%%%%%
%%%%%%%%%%%%%%%%%%%%%%%%%%%%%%%%%%%%%%%%%%%%%%%%%%%%%%%%%%%%%%%%%%%%%%
In this section we are going to present our numerical results. We organise the
discussion in the following way: we first look at the two operators we have
used separately. For each of them we examine the mass and volume dependence of
the numerical effort without and with improvements for the solvers and
preconditioning techniques switched on. For the overlap operator we then test
in addition the low mode preconditioning approach in the $\epsilon$--regime.
After discussing them separately we will then compare the two operators by
means of the best solver.

The algorithms are compared for each operator using the following criteria:
\begin{enumerate}
\item The \emph{total iteration number}:\\
  The number of iterations to reach convergence is a quantity which is
  independent of the detailed implementation of the Dirac operator as well as
  of the machine architecture, and therefore it provides a fair measure for
  comparison.

\item The \emph{total number of applications of $Q$}:\\
  In particular in case of the adaptive precision algorithms of the overlap
  operator, it turns out that the cost for one iteration depends strongly on
  the algorithm details, so a fairer mean for comparison in that case is the
  total number of applications of the Wilson-Dirac operator, i.e.~the number
  of $Q$ applications. Again this yields a comparative measure independent of
  the architecture and the details of the operator implementation, but on the
  other hand one should keep in mind that these first two criteria neglect the
  cost stemming from scalar products and ZAXPY operations. In particular this
  concerns the GMRES algorithm that needs significantly more of these
  operations than the other algorithms. It also concerns the adaptive
  precision algorithms for the overlap operator for reasons explained below.

\item The \emph{total execution time in seconds}:\\
  Finally, in order to study the relative cost factor between the inversion of
  the TM and the overlap operator we measure for each operator and algorithm
  the absolute timings on a specific machine, in our case on one node of the
  J{\"u}lich Multiprocessor (JUMP) IBM p690 Regatta using 32 processors.
  Obviously, these results will depend on the specific details of the machine
  architecture and the particular operator and linear algebra implementation,
  and hence will have no absolute validity.  Nevertheless, it is interesting
  to strive to such a comparison simply to obtain at least a feeling for the
  order of magnitude of the relative cost.

\end{enumerate}

%%%%%%%%%%%%%%%%%%%%%%%%%%%%%%%%%%%%%%%%%%%%%%%%%%%%%%%%%%%%%%%%%%%%%%
\subsection{Set-up}
\label{subsec:Setup}
%%%%%%%%%%%%%%%%%%%%%%%%%%%%%%%%%%%%%%%%%%%%%%%%%%%%%%%%%%%%%%%%%%%%%%
Our set-up consists of two quenched ensembles of 20 configurations with
volumes $V=12^4$ and $16^4$ generated with the Wilson gauge action at
$\beta=5.85$ corresponding to a lattice spacing of $a \sim 0.125$ fm
($r_0=0.5$ fm).

The bare quark masses for the overlap operator and the twisted mass operator
are chosen such that the corresponding pion mass values are matched, cf.~table
\ref{tab:matched quark masses}. Note that for the low mode preconditioning of
the overlap operator we consider an additional small mass which should bring
the system into the $\varepsilon$--regime.

\begin{table}[thb]
  \centering
    \begin{tabular*}{.5\textwidth}{@{\extracolsep{\fill}}ccl}
      \hline
      $\Bigl.\Bigr.m_\pi$[MeV] & $\mu_\textrm{ov}$ & $\mu_\textrm{tm}$ \\
      \hline\hline
      720 & 0.10  &   0.042   \\
      555 & 0.06  &   0.025   \\
      390 & 0.03  &   0.0125  \\
      230 & 0.01  &   0.004   \\
$\epsilon$-regime & 0.005 &    --      \\
      \hline
    \end{tabular*}
    \caption{Bare quark masses for the overlap and the twisted mass operator
      matched by the pion mass. The quark mass of $\mu_\textrm{ov}=0.005$
      corresponds to a simulation point in the $\epsilon$-regime, 
      where the notion of a pion mass becomes meaningless.}
  \label{tab:matched quark masses}
\end{table}

We invert the twisted mass (the overlap) operator on one (two) point-like
source(s) $\eta$ for each configuration at the four bare quark masses. The
required stopping criterion is $\|r\|^2=\|Ax-\eta\|^2 < 10^{-14}$, where $r$
is the residual and $x$ the solution vector. We are working in a chiral basis
and the two sources for the overlap operator are chosen such that they
correspond to sources in two different chiral sectors.  This is relevant for
the overlap operator only, which might have exact zero modes of the massless
operator in one of the two chiral sectors, potentially leading to a quite
different convergence behaviour. Furthermore the chiral sources allow to use
the chiral version of the CGNE algorithm for the overlap operator as described
in section \ref{sec:Lattice Dirac operators}. There it is also mentioned that
for the overlap operator we project out the lowest 20 and 40 eigenvectors of
$Q^2$ on the $12^4$ and $16^4$ lattice, respectively, in order to make the
construction of the $\sign$-function feasible.

For both operators we follow the strategy to first consider the
not preconditioned algorithms and then to switch on the available
preconditionings or improvements. Since for the overlap operator we have a
large range of algorithms to test (and the tests are more costly), we perform
the first step only at two masses and study the improvements from the
preconditioning and the full mass dependence only for a selection of
algorithms.

%%%%%%%%%%%%%%%%%%%%%%%%%%%%%%%%%%%%%%%%%%%%%%%%%%%%%%%%%%%%%%%%%%%%%%
\subsection{Twisted mass results}
%%%%%%%%%%%%%%%%%%%%%%%%%%%%%%%%%%%%%%%%%%%%%%%%%%%%%%%%%%%%%%%%%%%%%%
Before presenting results for the un-preconditioned TM Dirac operator, we need
to discuss the following point: the number of iterations needed by a certain
iterative solver depends in the case of the twisted mass Dirac operator
strongly on whether $D_\mathrm{tm}$ is inverted on a source $\eta$ or
$\gamma_5 D_\mathrm{tm}$ on a source $\gamma_5\eta$. This is due to the fact
that multiplying with $\gamma_5$ significantly changes the eigenvalue
distribution of the TM operator. All eigenvalues of $\gamma_5 D_\mathrm{tm}$
lie on a line parallel to the real axis shifted in the imaginary direction by
$\mu$, because the pure Wilson-Dirac operator obeys the property
$D_\mathrm{W}^\dagger=\gamma_5 D_\mathrm{W}\gamma_5$. To give examples, for
the BiCGstab and the GMRES algorithms $\gamma_5 D_\mathrm{tm}$ is
advantageous, while the CGS solver works better with $D_\mathrm{tm}$ itself.

This result is not surprising: it is well known that for instance the BiCGstab
iterative solver is not efficient, or even does not converge, when the
eigenvalue spectrum is complex and in exactly such situations the CGS
\cite{Sonneveld:1989cgs} algorithm often performs better.  Of course, for the
CG solver this question is not relevant, since in that case the operator
$D^\dagger D$ is used. Let us also mention that neither the MR nor the
MR$\gamma_5$ iterative solver converged for the twisted mass operator within a
reasonable number of iterations.

\begin{figure*}[htb]
  \begin{center}
    \begin{eqnarray*}
      \includegraphics[angle=0,width=10.0cm]{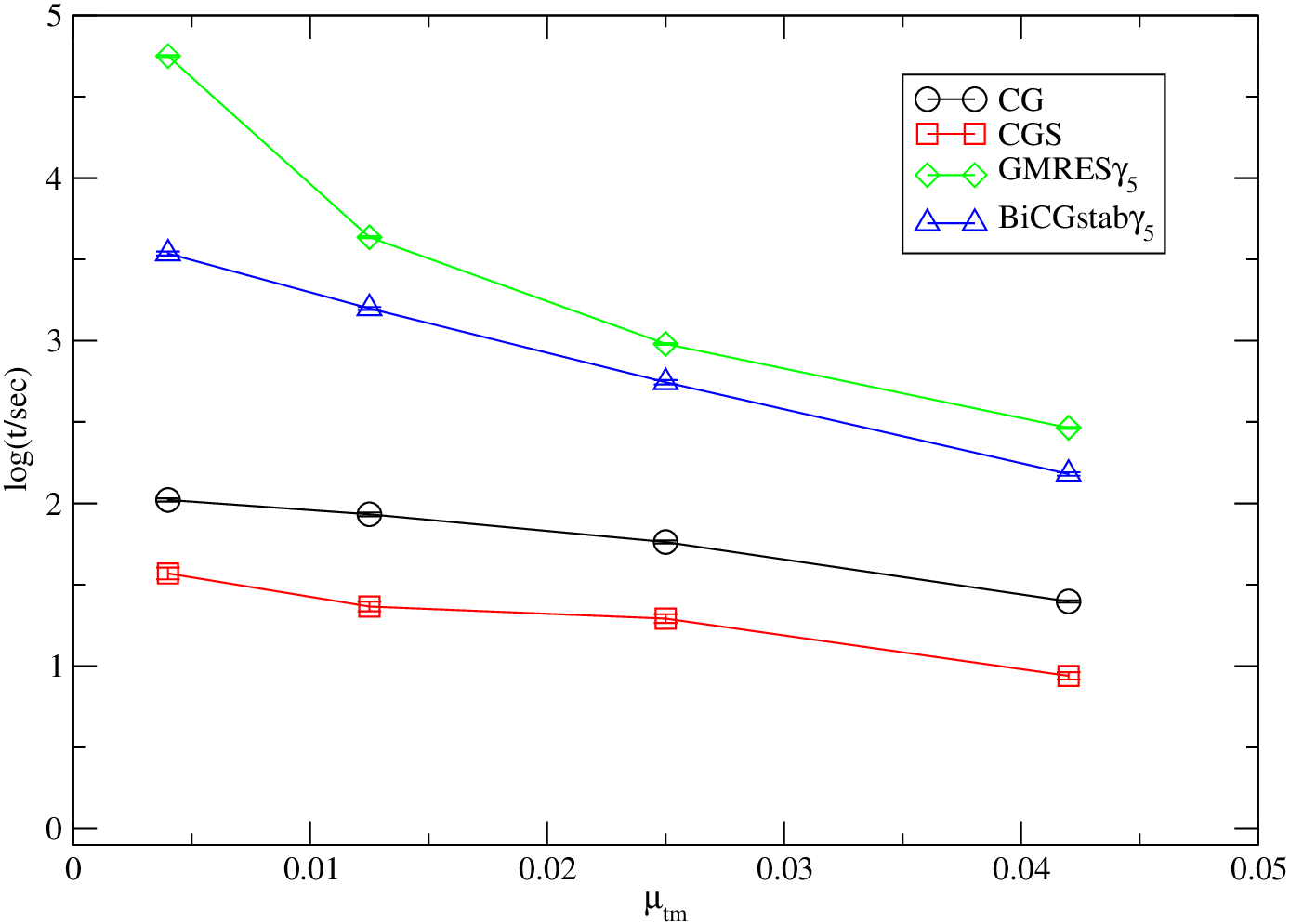} \\
      \includegraphics[angle=0,width=10.0cm]{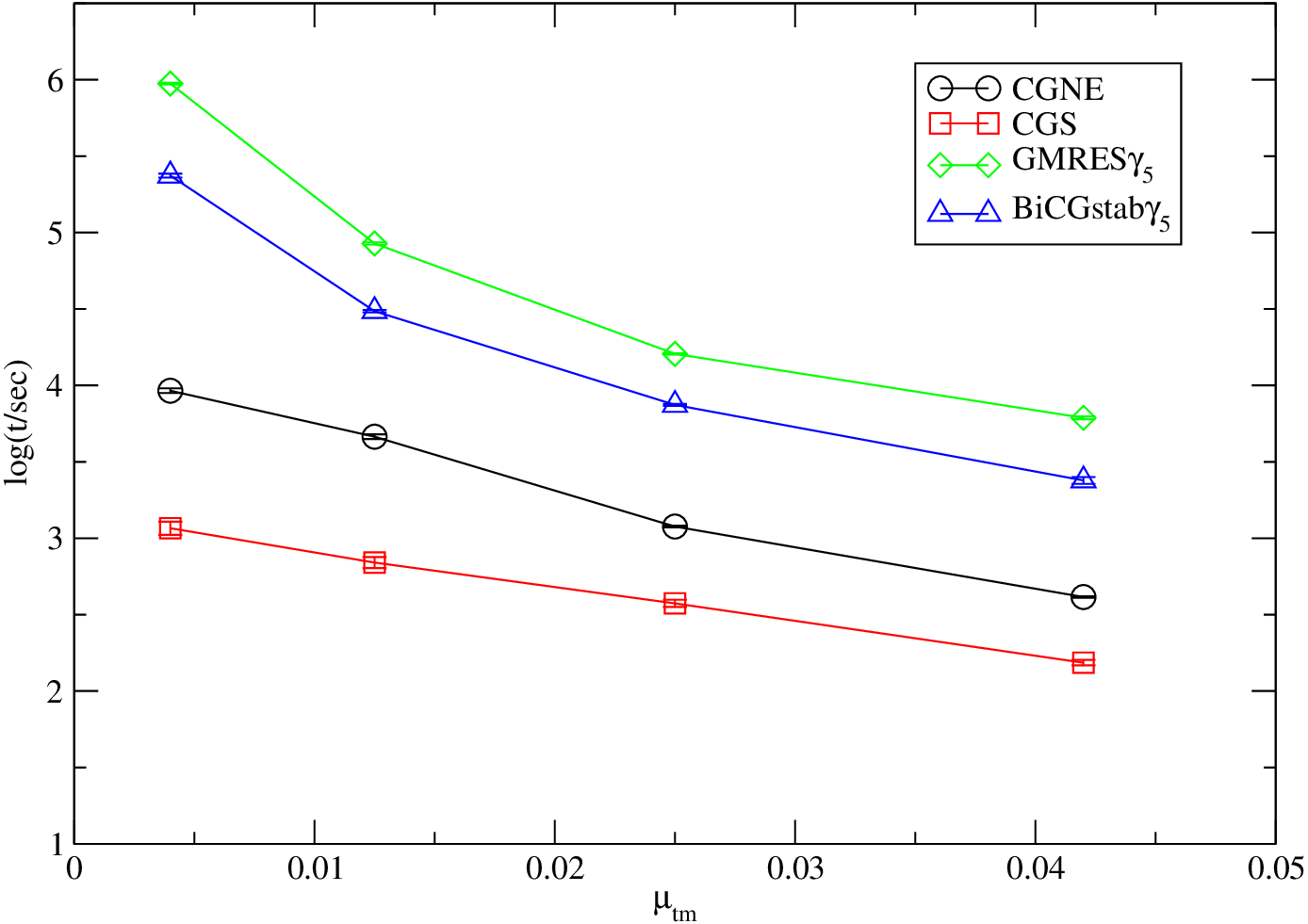}
    \end{eqnarray*}
  \end{center}
  \caption{Average timings for the inversion of the un-preconditioned Wilson TM
    operator in units of seconds on a logarithmic scale for
    different bare quark masses. We compare two volumes, a $12^4$ (top) and
    a $16^4$ lattice (bottom).}
  \label{fig:log_timings_tm}
\end{figure*}

The results for the un-preconditioned Wilson TM operator are collected in table
\ref{tab:TM number of iterations} where we give the average number of operator
applications (MV applications) that are required to reach convergence together
with the standard deviation. In the case of the TM operator, the number of MV
applications is proportional to the number of solver iterations where the
proportionality factor can be read off column 4 in table \ref{tab:alglist}.
\begin{table}[thb]
  \begin{center}
    \begin{tabular*}{1.\textwidth}{@{\extracolsep{\fill}}lcccc}
      \hline\hline
      & $\mu_\textrm{tm}=0.042$& $0.025$ & $0.0125$  & $0.004$  \\
      \hline\hline
      $V=12^4$ & & & & \\
      \hline
      CGNE               & 2082(60)  & 2952(175) & 3536(234) & 3810(243)  \\
      CGS                & 1251(178) & 1661(262) & 1920(361) & 2251(553)  \\
      BiCGstab$\gamma_5$ & 3541(175) & 5712(280) & 9764(503) & 12772(979)  \\
      GMRES$\gamma_5$    & 1962(48)  & 3314(92)  & 6223(199) & 19204(737)  \\
      \hline
      $V=16^4$ & & & & \\
      \hline
      CGNE               & 2178(46)  & 3556(107) & 6277(414) & 8697(802)  \\
      CGS                & 1336(134) & 2029(276) & 2614(508) & 3420(866)  \\
      BiCGstab$\gamma_5$ & 3526(145) & 5805(239) & 10940(547)& 26173(2099)  \\
      GMRES$\gamma_5$    & 1945(42)  & 3287(78)  & 6168(129) & 19106(565)  \\
      \hline\hline
    \end{tabular*}\\
    \caption{Average number (and standard deviation) of MV applications for
      reaching convergence of the un-preconditioned Wilson TM operators. 
      Here and in the following tables, averages are always taken over 
      20 independent pure gauge configurations.}
    \label{tab:TM number of iterations}
  \end{center}
\end{table}
From these data it is clear that the CGS algorithm is the winner for all
masses and on both volumes. The CGS algorithm shows a rather weak exponential
mass dependence and beats the next best algorithm CGNE by a factor 2.5 at the
smallest mass on the large volume as is evident from figure
\ref{fig:log_timings_tm} where we plot the logarithm of the absolute timings
in units of seconds as a function of the bare quark mass. Since the CGNE shows
a similar scaling with the mass as the CGS we do not expect this conclusion to
change for smaller masses. Moreover the CGS appears to have a weaker volume
dependence than the CGNE, in particular at small masses, so we expect the
conclusion to be strengthened as the volume is further increased.  A very
interesting point to note is that the GMRES$\gamma_5$ algorithm shows a
perfect scaling with the volume in the sense that the iteration numbers remain
constant as the volume is increased.

%%%%%%%%%%%%%%%%%%%%%%%%%%%%%%%%%%%%%%%%%%%%%%%%%%%%%%%%%%%%%%%%%%%%%%
\subsubsection{Even/odd preconditioning}
\label{subsubsec:Evenodd}
%%%%%%%%%%%%%%%%%%%%%%%%%%%%%%%%%%%%%%%%%%%%%%%%%%%%%%%%%%%%%%%%%%%%%%
Let us now present the results with even/odd preconditioning. For the
CGS$_\text{eo}$, BiCGstab$_\text{eo}$ and GMRES$_\text{eo}$ solvers (and their
$\gamma_5$ versions) we used the symmetric even/odd preconditioning as
outlined at the end of section \ref{subsec:eo}, while for the CGNE we used the
non-symmetric version. The results for the average number of operator
applications required to reach convergence together with the standard
deviation are collected in table \ref{tab:tmeoiter}.

As in the case of the un-preconditioned operator also with even/odd
preconditioning it makes a difference whether the $\gamma_5$ version of a
solver is used or not. We will discuss these differences here in more detail.
The GMRES$_\text{eo}$ solver for instance stagnates on most of the 20
configurations for both lattice sizes, while the GMRES$\gamma_5$$_\text{eo}$
converges without problems. The BiCGstab$_\text{eo}$ algorithm on the other
hand does not converge on one $12^4$ configuration and on six $16^4$
configurations, while again the BiCGstab$\gamma_5$$_\text{eo}$ algorithm
converges without any problem. In case the BiCGstab$_\text{eo}$ converges it
is much faster than the BiCGstab$\gamma_5$$_\text{eo}$, as can be seen in
table \ref{tab:tmeoiter}. On the other hand the iteration numbers of
BiCGstab$_\text{eo}$ for the larger volume show only a very weak mass
dependence and the variance is large. This might indicate that the number of
configurations where the BiCGstab$_\text{eo}$ does not converge is likely to
increase further, if the volume is increased.

A similar picture can be drawn for the CGS$_\text{eo}$ and
CGS$\gamma_5$$_\text{eo}$ solvers, but in this case the CGS$_\text{eo}$
converges in all cases and is moreover the fastest algorithm for both lattice
sizes and all masses.

\begin{table}[thb]
  \begin{center}
    \begin{tabular*}{1.\textwidth}{@{\extracolsep{\fill}}lcccc}
      \hline\hline
      & $\mu_\textrm{tm}=0.042$& $0.025$ & $0.0125$  & $0.004$  \\
      \hline
      $V=12^4$ & & & & \\
      \hline
      CGNE$_\text{eo}$               & 725(18) & 1042(64) & 1238(91) & 1333(93)  \\
      CGS$\gamma_5$$_\text{eo}$      & 2999(269) & 2788(265) & 2659(212) & 2526(198)  \\
      CGS$_\text{eo}$                & 599(87) & 774(135) & 944(169) & 1048(234)  \\
      BiCGstab$\gamma_5$$_\text{eo}$ & 1279(64) & 2060(123) & 3353(189) & 4103(382)  \\
      BiCGstab$_\text{eo}$           & 799(293) & 880(337) & 1545(1607) & 2044(2801)  \\
      GMRES$\gamma_5$$_\text{eo}$    & 731(19) & 1180(35) & 2261(75) & 6670(258)  \\
      \hline
      \hline
      $V=16^4$ & & & & \\
      \hline
      CGNE$_\text{eo}$               & 755(14) & 1227(37) & 2187(147) & 3048(289)  \\
      CGS$\gamma_5$$_\text{eo}$      & 10408(2043) & 8332(1399) & 7014(581) & 6819(1491)  \\
      CGS$_\text{eo}$                & 650(60) & 962(151) & 1317(252) & 1687(448)  \\
      BiCGstab$\gamma_5$$_\text{eo}$ & 1290(71) & 2063(94) & 3892(183) & 8786(730)  \\
      BiCGstab$_\text{eo}$           & 1595(595) & 1705(928) & 1576(868) & 1884(1501)  \\
      GMRES$\gamma_5$$_\text{eo}$    & 728(13) & 1174(21) & 2258(42) & 6722(145)  \\
      \hline\hline
    \end{tabular*}
    \caption{Average number (and variance) of MV applications for
      convergence of the even/odd preconditioned  Wilson TM
      operators.} 
    \label{tab:tmeoiter}
  \end{center}
\end{table}

\begin{figure*}[thb]
  \begin{center}
    \begin{eqnarray*}
      \includegraphics[angle=0,width=10.0cm]{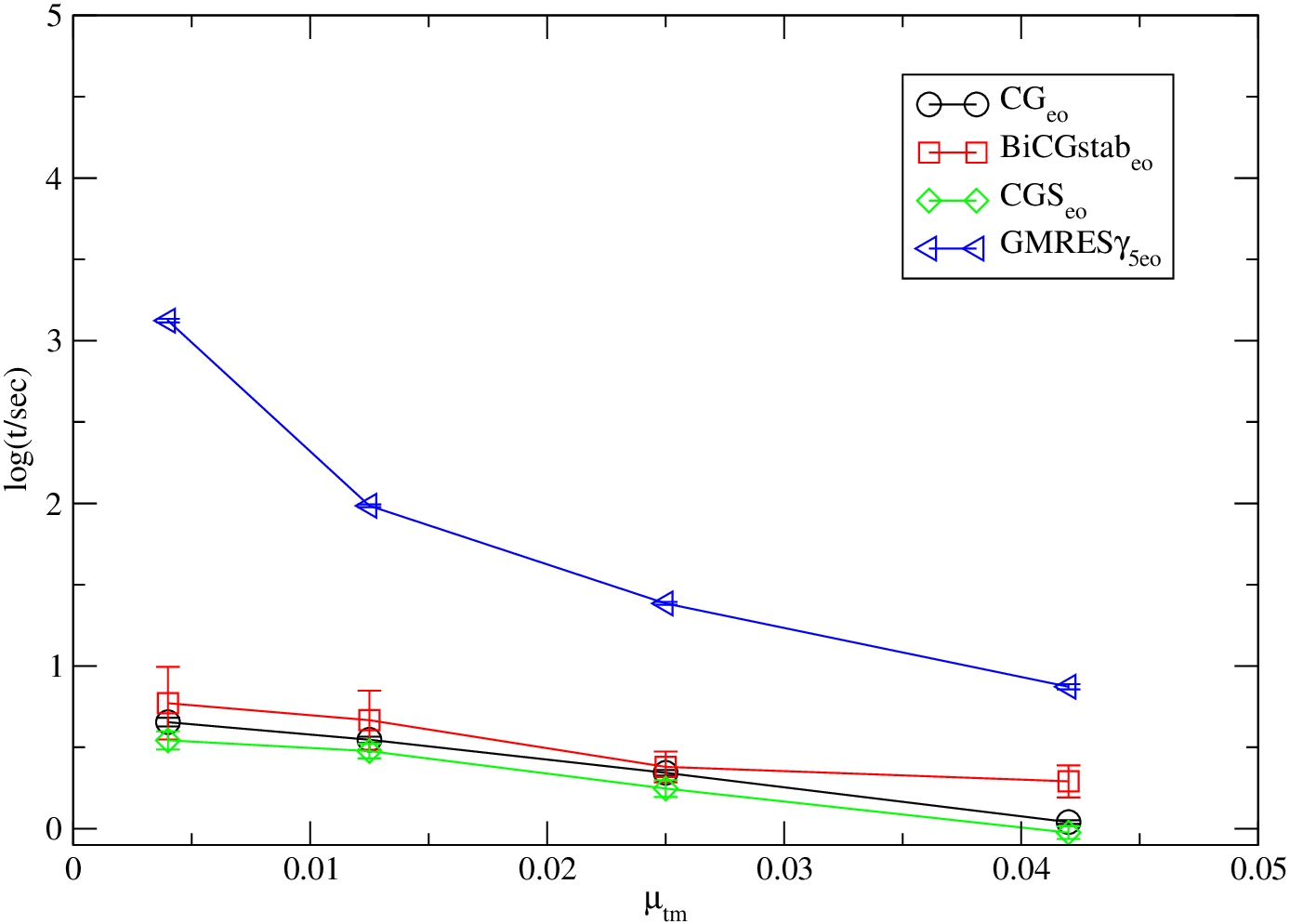} \\
      \includegraphics[angle=0,width=10.0cm]{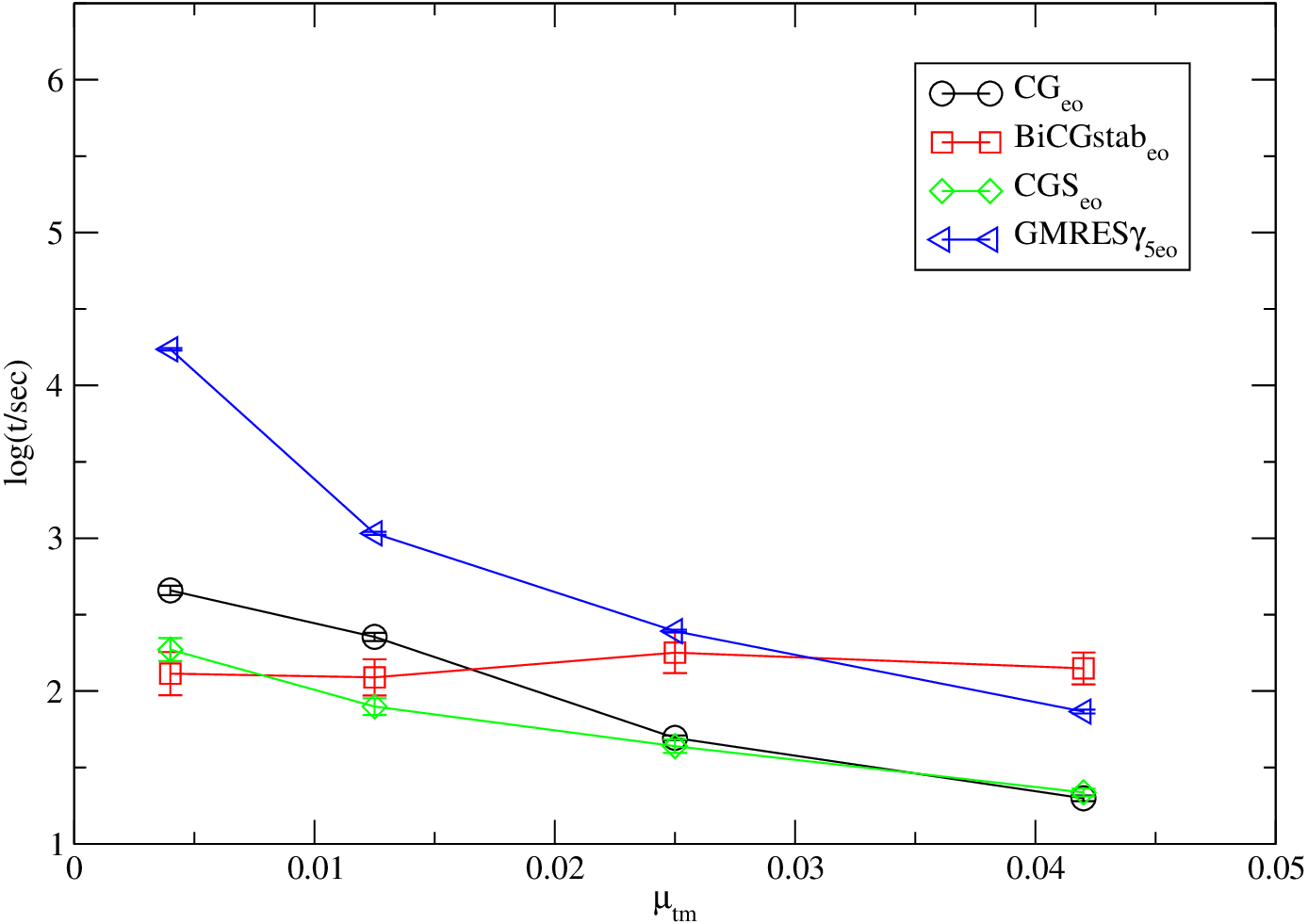}
    \end{eqnarray*}
  \end{center}
  \caption{{}Average timings for the inversion of the even/odd 
    preconditioned Wilson TM
    operator in units of seconds on a logarithmic scale for
    different bare quark masses. We compare two volumes, a $12^4$ (top) and
    the $16^4$ lattice (bottom).}
  \label{fig:log_timings_tm_eo}
\end{figure*}

The next to best algorithms are the CGNE and BiCGstab$_\text{eo}$, where the
latter has the drawback of non-convergence and instabilities for a certain
number of configurations.  Therefore, concentrating on the CGNE and the
CGS$_\text{eo}$, we observe that in particular on the larger volume the
CGS$_\text{eo}$ shows a better scaling with the mass: while the
CGS$_\text{eo}$ is at the largest mass only a factor 1.16 faster, this factor
increases to 1.8 at the smallest mass value. At this point a comparison in
execution time is of interest, cf.~fig.\ref{fig:log_timings_tm_eo} , because
the number of SP and ZAXPY operations for each iteration are different for the
various solvers. We find that CGS$_\text{eo}$ remains the most competitive
algorithm given the fact that BiCGstab$_\text{eo}$ is not always stable.
On the other hand the situation could change in favour of the
GMRES$_\text{eo}$ algorithm for large volumes, since the CGS$_\text{eo}$ has a
much worse volume dependence than the GMRES$_\text{eo}$ which again shows a
perfect scaling with the volume like in the un-preconditioned case.

Finally we note that comparing the best algorithm for the even/odd
preconditioned operator to the one for the un-preconditioned operator we
observe a speed-up of about 2 for our investigated range of parameters.
%\clearpage

%%%%%%%%%%%%%%%%%%%%%%%%%%%%%%%%%%%%%%%%%%%%%%%%%%%%%%%%%%%%%%%%%%%%%%
\subsection{Overlap results}
%%%%%%%%%%%%%%%%%%%%%%%%%%%%%%%%%%%%%%%%%%%%%%%%%%%%%%%%%%%%%%%%%%%%%%
Let us first have a look at the results of the overlap operator without any
improvements or preconditioning. As noted in the introduction to this section
we have investigated the full mass scaling of the un-preconditioned algorithms
only for a selection of algorithms, in particular we have done this for the
adaptive precision versions to be discussed later.  The results are collected
in table \ref{tab:overlap number of iterations} where we give the average
number of operator applications (MV applications) that are required to reach
convergence together with the standard deviation. We note again that the
number of MV applications is proportional to the number of iterations where
the proportionality factor can be read from column 4 in table
\ref{tab:alglist}.
\begin{table}[thb]
  \begin{center}
    \begin{tabular*}{1.\textwidth}{@{\extracolsep{\fill}}lcccc}
      \hline\hline
      & $\mu_\textrm{ov}=0.10$& $0.06$ & $0.03$  & $0.01$  \\
      \hline
      $V=12^4$ & & & &   \\
      \hline
      CGS              & 239(22) &   --  & 593(88) &   --   \\
      BiCGstab         & 207(13) & 333(24) & 549(55) & 695(108)  \\
      MR               & 206( 3) &   --  & 646(16) &   --   \\
      % mr$_\textrm{ap}$ & 231( 8) & 373(14) & 759(45) & 2647(383)  \\
      GMRES            & 187( 6)  &   --  & 576(37) &   --   \\
      % gmres$_\textrm{ap}$ & 212( 9) & 326(14) & 616(43) & 1749(262)  \\
      SUMR             & 174( 7)   & 260(19) & 350(46) & 394(55)  \\
      % sumr$_\textrm{ap}$ & 180( 7) & 267(20) & 354(46) & 397(55)  \\
      CG               & 336(33) &   --  & 411(52) &   --   \\
      CG$_\chi$        & 168(17) &   --  & 205(26) &   --   \\
      % cg$_{\chi,\textrm{ap}}$ & 168(17) & 193(22) & 205(26) & 234(47)  \\
      \hline\hline
      $V=16^4$ & & & &   \\
      \hline
      CGS              & 241(19) &   --  & 738(71) &   --   \\
      BiCGstab         & 212(10) & 340(17) & 647(36) & 1552(215)  \\
      MR               & 206( 3) &   --  & 644(14) &   --   \\
      % mr$_\textrm{ap}$ & 229( 5) & 366(11) & 719(20) & 2217(85)  \\
      GMRES            & 187( 5) &   --  & 584(19) &   --   \\
      % gmres$_\textrm{ap}$ & 215( 5) & 331( 9) & 633(21) & 1899(99)  \\
      SUMR             & 179( 5) & 284( 9) & 523(26) & 929(124)  \\
      % sumr$_\textrm{ap}$ & 184( 5) & 293( 9) & 539(27) & 937(121)  \\
      CG               & 411(11) &   --  & 949(105) &   --   \\
      CG$_\chi$        & 206( 6) &   --  & 475(52) &   --   \\
      % cg$_{\chi,\textrm{ap}}$ & 208( 6) & 336(14) & 478(52) & 543(89)  \\
      \hline\hline
    \end{tabular*}
    \caption{Average number (and standard deviation) of MV applications for
      convergence of the overlap operator.}
    \label{tab:overlap number of iterations}
  \end{center}
\end{table}
The first thing we note is that the iteration numbers are much smaller than
for the Wilson TM operator, usually by about one order of magnitude. This is
presumably due to the fact that the spectrum of the overlap operator is much
more restricted to lie exactly on the Ginsparg-Wilson circle and better
behaved than the one of the TM operator, and usually iterative inversion
algorithms are very sensitive to the distribution of the eigenvalues.

From the results in table \ref{tab:overlap number of iterations} we do not
find a completely coherent picture, but we may say that at least at small
quark mass CG$_\chi$ is the winner followed by SUMR and GMRES. Looking at the
mass scaling behaviour it appears that CG$_\chi$ shows the weakest dependence
on the mass and so this conclusion should hold towards smaller quark masses.
Concerning the volume dependence we note that at the smallest mass the
CG$_\chi$ and SUMR have a very similar behaviour and so again the conclusion
should not be changed at larger volumes. However, as for the Wilson TM
operator the GMRES algorithm, and in addition here also the MR, shows a
perfect scaling behaviour with the volume. Towards small quark masses this
positive finding is compensated by the bad scaling of these two algorithms
with the mass, but for intermediate quark masses we can expect both GMRES and
MR to be superior to the SUMR and CG$_\chi$, at least on large enough volumes.

\begin{figure*}[htb]
  \begin{center}
    \begin{eqnarray*}
      \includegraphics[angle=0,width=10.0cm]{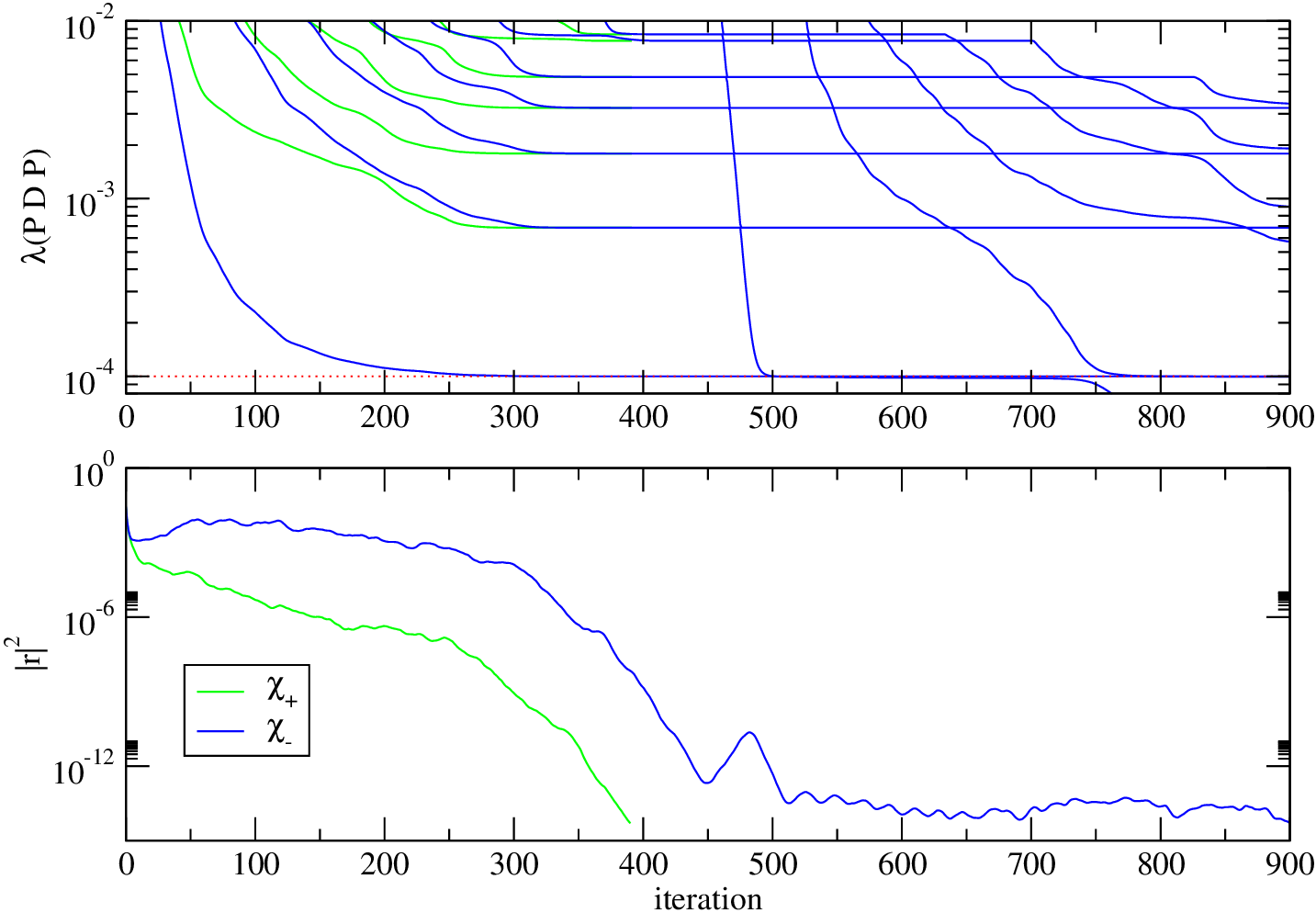} 
    \end{eqnarray*}
  \end{center}
  \caption{{}Convergence history of CG$_\chi$ on the 'worst case' $16^4$ 
    configuration at $\beta=5.85$ and $\mu=0.01$. The lower plot shows the
    iterated residual while the upper plot shows the eigenvalues of the
    corresponding underlying Lanczos iteration matrix encountering spurious
    zero modes.}
  \label{fig:bad_convergence}
\end{figure*}

Let us finally make a cautionary remark on the CG$_\chi$ algorithm. It is
clear that Eq.(\ref{eq:PDP operator}) holds only for the exact overlap
operator and any approximation to it will introduce some corrections. Indeed,
the approximation errors on both sides of Eq.(\ref{eq:PDP operator}) are
rather different. If we assume a maximal error $\delta$ over the interval of
our approximation to the $\sign$ function, then the l.h.s.~has an error
bounded by $(1-\mu)\delta |D|$ while for the r.h.s.~it is
$(1-\mu^2/2M)\delta$. As a consequence the two operators do agree only up to a
certain accuracy level and the agreement deteriorates towards small quark
masses where the lowest modes of $D$ become important. E.g.~in propagator
calculations this is reflected in the fact that a solution calculated with one
operator to some accuracy is in fact not a solution of the other operator to
the same accuracy. In practise we have observed this phenomenon only at the
smallest quark mass $\mu=0.01$ and mainly on the $16^4$ lattices where we
found accuracy losses in the true residuals of up to two orders of magnitude,
i.e.~$|r|^2 < 10^{-14}$ versus $|r|^2 < 10^{-12}$, even though our
approximations of $D$ satisfy the Ginsparg-Wilson relation to machine
precision. Moreover, in those cases we have usually observed a rather strange
convergence behaviour which can be related to the occurrence of spurious zero
modes in the underlying Lanczos iteration matrix. As an illustration we show
in figure \ref{fig:bad_convergence} the worst case that we encountered. In the
lower plot we show the iterated residual as a function of the iteration number
while in the upper plot we show the eigenvalues of the corresponding
underlying Lanczos iteration matrix (cf.~appendix \ref{subsec:indexCG} for
additional explanations).

One possible remedy for all this is to simply stop the CG$_\chi$ algorithm
shortly before this happens, e.g.~in the above case as soon as the iterated
residual reaches $|r|^2 < 10^{-12}$, and to restart with the standard CG
algorithm. Convergence is then usually reached within a small number of
iterations, but obviously the MM capability is lost.

%\clearpage

%%%%%%%%%%%%%%%%%%%%%%%%%%%%%%%%%%%%%%%%%%%%%%%%%%%%%%%%%%%%%%%%%%%%%%
\subsubsection{Adaptive precision}
\label{subsubsec:overlap ap results}
%%%%%%%%%%%%%%%%%%%%%%%%%%%%%%%%%%%%%%%%%%%%%%%%%%%%%%%%%%%%%%%%%%%%%%
Let us now turn to the adaptive precision algorithms for the overlap operator.
As noted before we have implemented the adaptive precisions for the MR,
GMRES, SUMR and CG$_\chi$ algorithms. Without quoting the numbers we remark
that the iteration numbers (at the parameter points where we can compare) for
the CG$_{\chi,\textrm{ap}}$ and the SUMR$_\textrm{ap}$ are the same as for the
corresponding algorithms without adaptive precision (within 0-3\%), while for
the other two, MR$_\textrm{ap}$ and GMRES$_\textrm{ap}$, the iteration numbers
increase by about 7-15\%. This can be understood by the fact that the latter
two algorithms involve several correction steps with subsequent restarts as
explained in section \ref{subsec:adaptive precision} therefore undermining
slightly the efficiency of the algorithms.

\begin{figure*}[bth]
  \begin{center}
    \vspace{-1cm}
    \begin{eqnarray*}
      \includegraphics[angle=0,width=10.0cm]{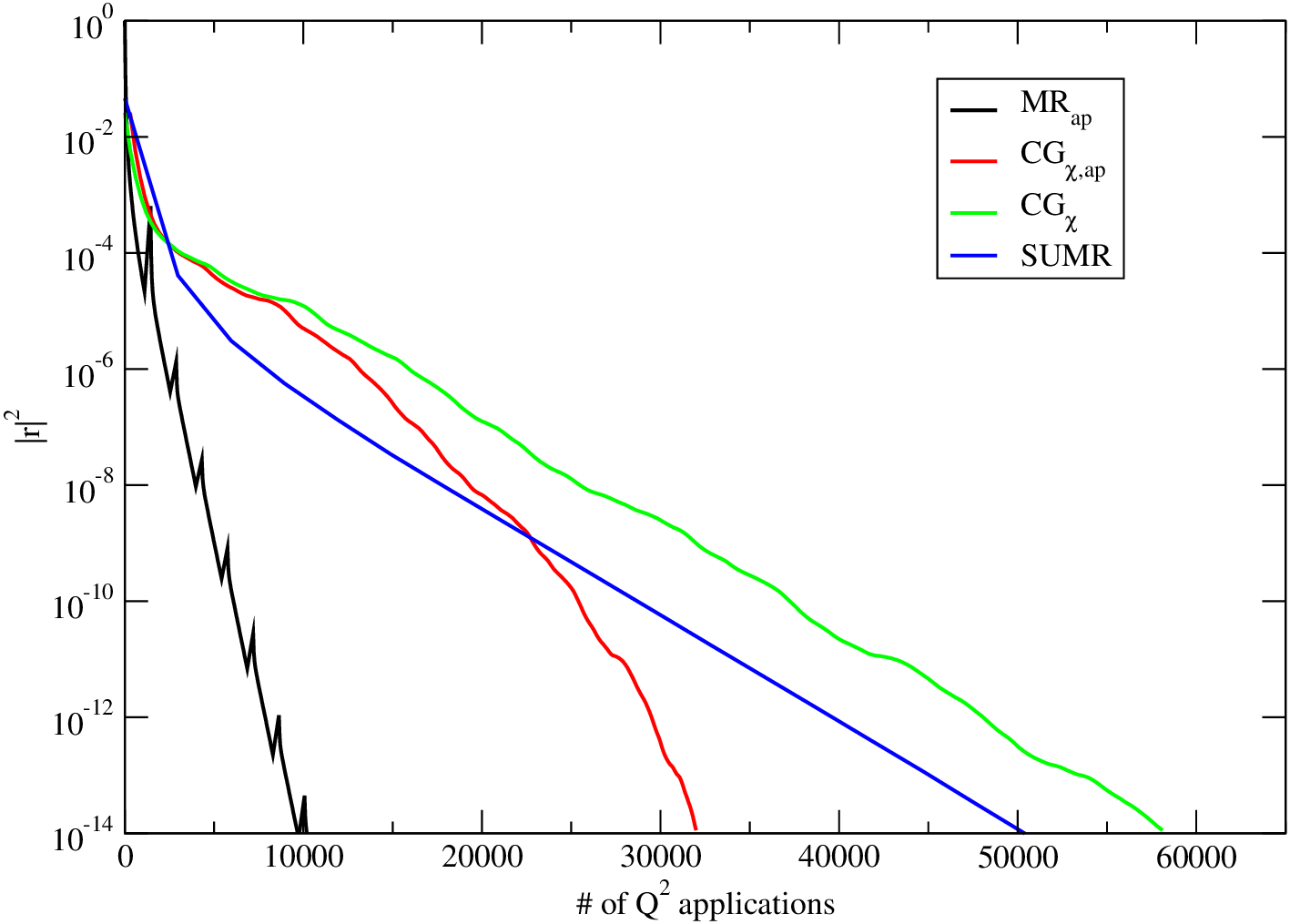} \\
      \includegraphics[angle=0,width=10.0cm]{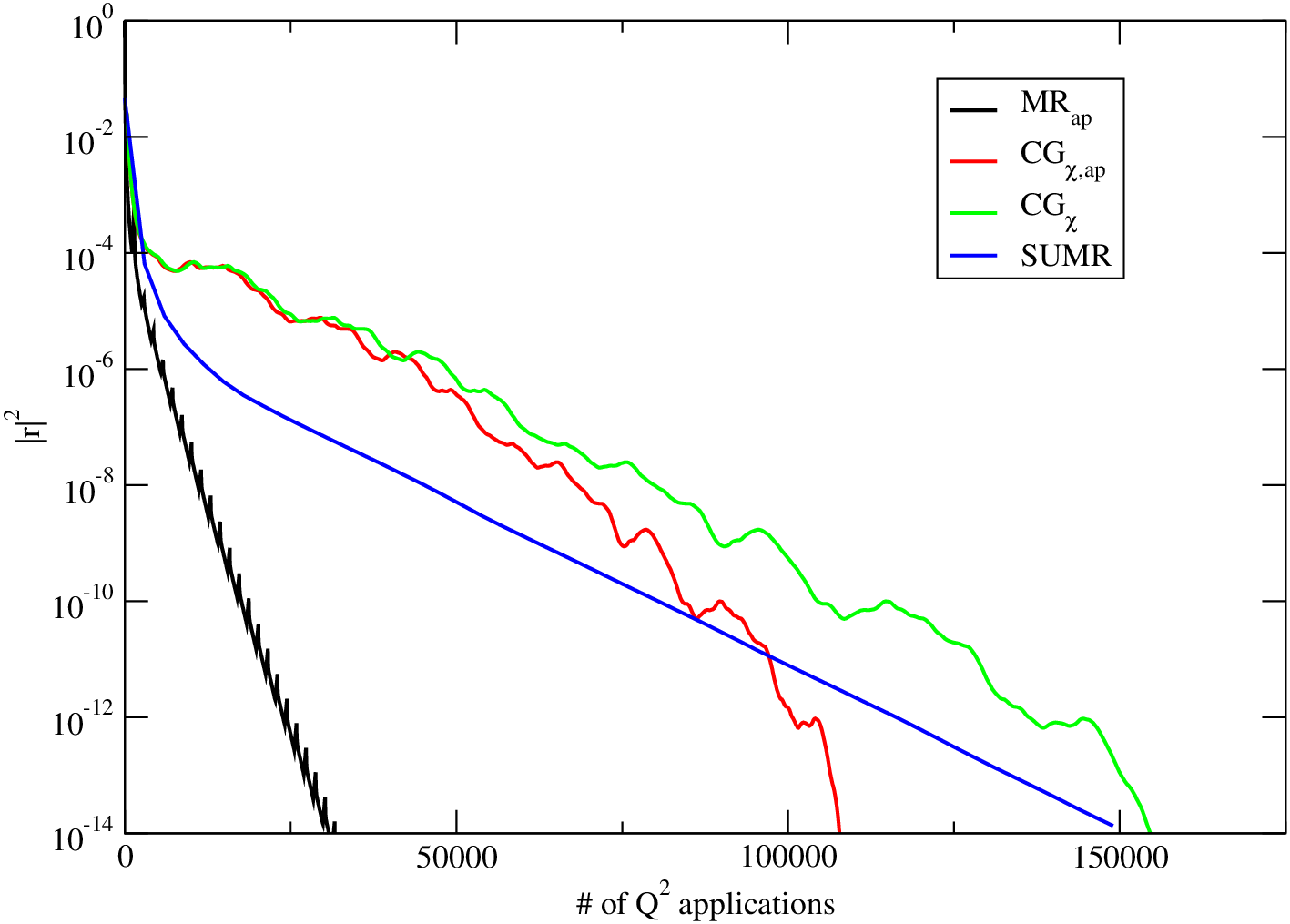}
    \end{eqnarray*}
  \end{center}
  \caption{{}Convergence history of SUMR, CG$_\chi$ compared to adaptive
    precision CG$_{\chi,\text{ap}}$ and MR$_\text{ap}$ for the overlap operator
    on the $16^4$ lattice at $\beta=5.85$ with $\mu=0.10$ (top) and
    $\mu=0.03$ (bottom).}  
  \label{fig:ap}
\end{figure*}

However, it should be clear from section \ref{subsec:adaptive
  precision} that the iteration number is not 
the crucial quantity here, but instead it is the total number of applications
of the Wilson kernel, i.e.~$Q$. This is exemplified in figure \ref{fig:ap}
where we show, in units of $Q$ applications, the convergence history of SUMR
and CG compared to CG$_\text{ap}$ and MR$_\text{ap}$ for the overlap operator
on the $16^4$ lattice at $\beta=5.85$ with $\mu=0.10$ (top) and $\mu=0.03$
(bottom).
\begin{table}[tb]
  \begin{center}
    \begin{tabular*}{1.\textwidth}{@{\extracolsep{\fill}}lcccc}
      \hline\hline
      & $\mu_\textrm{ov}=0.10$  & $0.06$     & $0.03$       & $0.01$  \\
      \hline
      $V=12^4$ & & & &   \\
      \hline
      MR                    & 103.2(10.2) & --          & 323.2(33.0) & --          \\
      MR$_\text{ap}$        & 18.5(1.9)   & 30.0(2.6)   & 61.1(4.3)   & 212.0(17.5) \\
      CG$_\chi$             & 84.2(13.5)  & --          & 103.1(18.9) & --          \\
      CG$_{\chi,\text{ap}}$ & 51.2(7.7)   & 73.2(13.0)  & 83.3(17.3)  & 96.7(22.6)  \\
      GMRES                 & 93.6(10.3)  & --          & 288.1(37.5) & --          \\
      GMRES$_\text{ap}$     & 18.1(2.1)   & 27.9(3.2)   & 52.9(7.0)   & 150.8(29.9) \\
      SUMR                  & 87.3(10.0)  & 130.5(18.9) & 175.7(33.9) & 198.0(39.8) \\
      SUMR$_\text{ap}$      & 55.8(6.7)   & 83.1(11.9)  & 118.7(22.3) & 146.7(31.8) \\
      \hline
      $V=16^4$ & & & & \\
      \hline
      MR                    & 126.2(9.6)  & --          & 394.2(29.8) & --          \\
      MR$_\text{ap}$        & 22.3(1.7)   & 35.8(2.7)   & 70.6(5.2)   & 218.2(14.9) \\
      CG$_\chi$             & 125.8(10.1) & --          & 291.4(44.1) & --          \\
      CG$_{\chi,\text{ap}}$ & 77.0(9.1)   & 134.8(11.2) & 215.3(31.7) & 281.1(54.6) \\
      GMRES                 & 114.7(8.9)  & --          & 357.8(29.0) & --          \\
      GMRES$_\text{ap}$     & 22.3(1.8)   & 34.5(2.7)   & 66.0(5.6)   & 198.5(18.2) \\
      SUMR                  & 109.4(8.6)  & 174.2(14.5) & 320.2(30.4) & 570.5(96.8) \\
      SUMR$_\text{ap}$      & 69.3(5.6)   & 108.5(9.2)  & 196.0(19.3) & 372.3(63.1) \\
      \hline\hline
    \end{tabular*}
    \caption{Average number (and standard deviation) of $Q$
      applications for convergence of the overlap operators, in units of
      1000. 
} 
    \label{tab:overlap number of Q applications}
  \end{center}
\end{table}
In table \ref{tab:overlap number of Q applications} we give the total number
of applications of the Wilson-Dirac operator $Q$ which again yields a measure
independent of the architecture and the details of the operator implementation
for a comparison among the algorithms. We find that the gain from the adaptive
precision for MR and GMRES is around a factor of 5.5, while it is around 1.5
for CG and SUMR. The gain deteriorates minimally towards smaller quark masses,
except for GMRES$_\text{ap}$ where it improves slightly. The difference of the
factors for the two sets of algorithms becomes evident by reflecting the fact
that the former use low order polynomials right from the start of the
algorithm while for the latter the adaptive precision becomes effective only
towards the end. Comparing among the algorithms we find that except for the
smallest mass on the smaller volume the best algorithm is GMRES$_\text{ap}$
almost matched by MR$_\text{ap}$. They are by a factor 2-3 more efficient than
the next best CG$_{\chi,\text{ap}}$ on the small volume and SUMR$_\text{ap}$
on the large one. This pattern can be understood by the bad scaling properties
of MR and GMRES, as opposed to CG and SUMR, towards small quark masses which
on the other hand is compensated at the larger volume by their almost perfect
scaling with the volume.

However, as discussed before this is not the whole story -- for a relative
cost estimate one has to keep in mind that each application of the
$\sign$-function, independent of the order of the polynomial for the
$\sign$-function approximation, generically requires the projection of $O(10)$
eigenvectors of $Q$ and this contributes a significant amount to the total
cost. This is particularly significant in the case of the MR$_\text{ap}$ and
GMRES$_\text{ap}$ both of which use low order approximations of the
$\sign$-function but require a rather large number of iterations (and
therefore many projections), so the total cost depends on a subtle interplay
between the number of scalar products (proportional to the number of
iterations in table \ref{tab:overlap number of iterations}) and the number of
$Q$ applications in table \ref{tab:overlap number of Q applications}.

\begin{figure}[htb]
  \centering
  \includegraphics[angle=0,width=10.0cm]{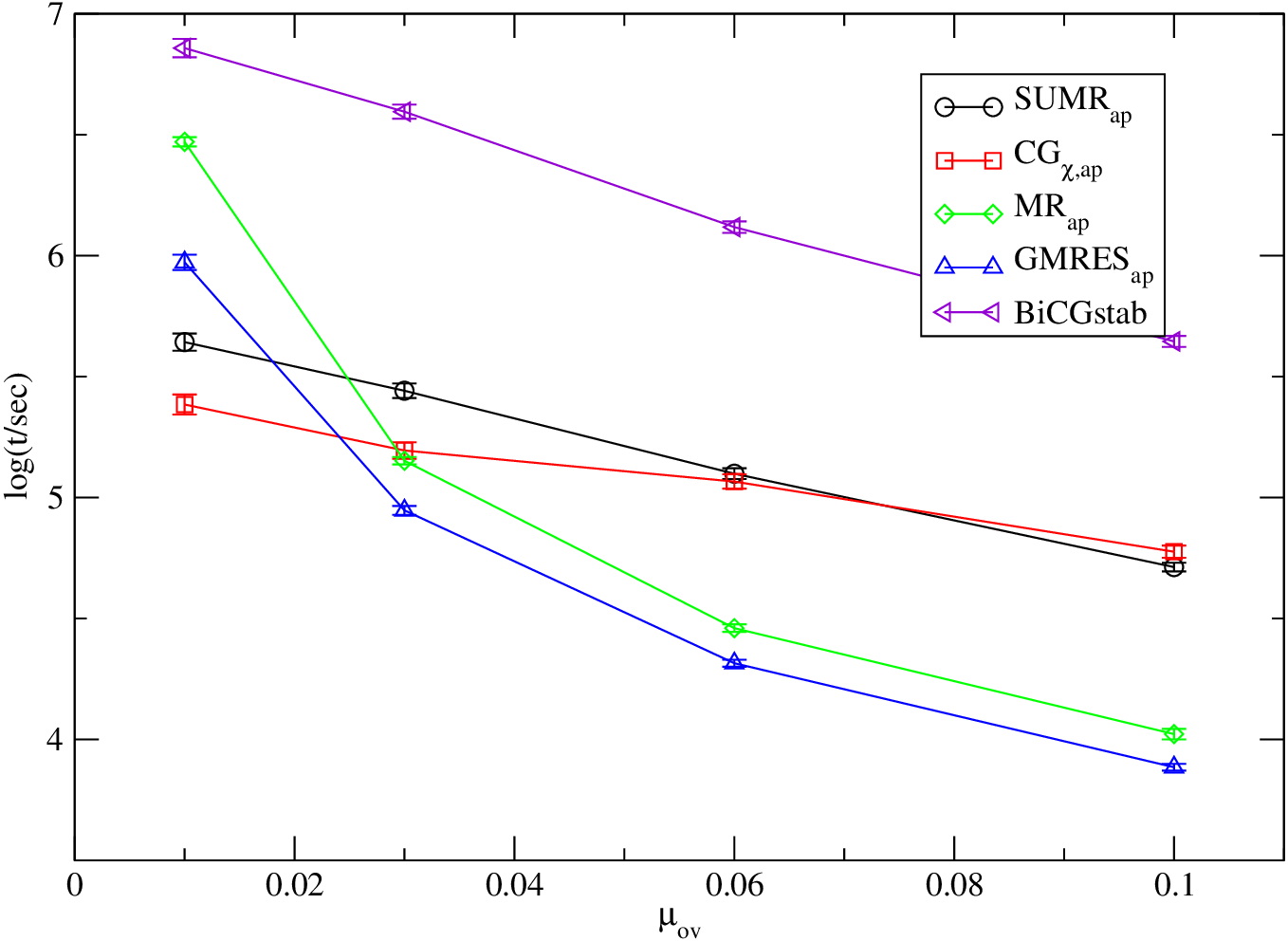} \\
  \includegraphics[angle=0,width=10.0cm]{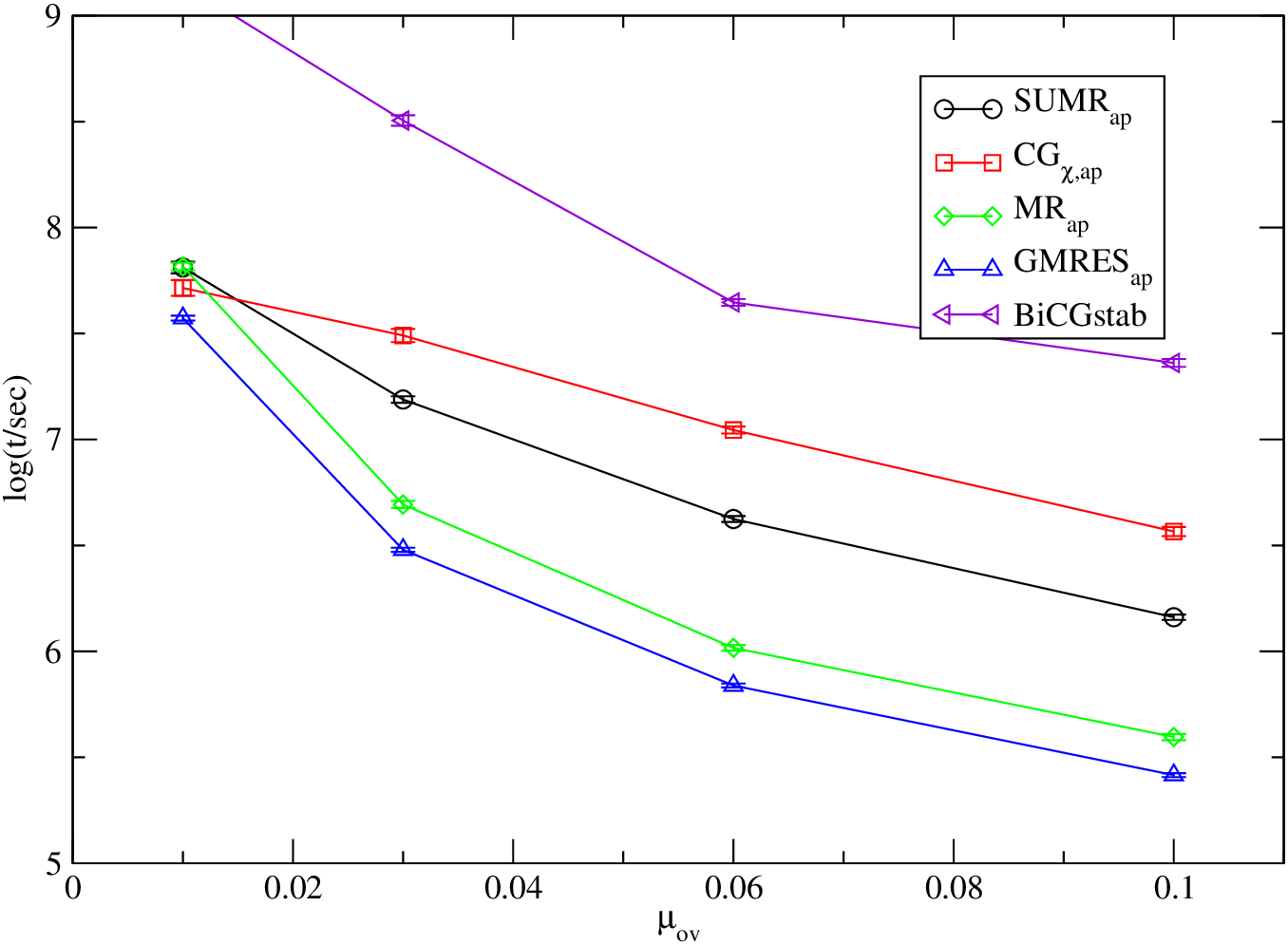} 
  \caption{Average timings for the inversion of the overlap
    operator in units of seconds on a logarithmic scale for
    different bare quark masses. At $\beta=5.85$ on the $12^4$ (top) and
    the $16^4$ lattice (bottom).}
  \label{fig:log_timings_overlap}
\end{figure}

In order to take this into account let us compare the absolute timings for the
adaptive precision algorithms. As emphasised before the results will obviously
depend on the specific MV, SP and ZAXPY implementation details as well as on
the architecture of the machine. In figure \ref{fig:log_timings_overlap} we
plot the logarithm of the absolute timings in units of seconds as a function of
the bare quark mass.

We note that on the more relevant larger volume the pattern follows
essentially the one observed for the numbers in table \ref{tab:overlap number
  of Q applications}. As before, GMRES$_\text{ap}$ and MR$_\text{ap}$ appear
to be more efficient than CG$_{\chi,\text{ap}}$ and SUMR$_\text{ap}$ except
for very small quark masses. However, the almost perfect volume scaling of
GMRES$_\text{ap}$ (and similarly MR$_\text{ap}$) suggests that these
algorithms will break even also at small masses on large enough volumes.
Indeed, as is evident from figure \ref{fig:log_timings_overlap}, this appears
to be the case already on the $16^4$ lattice where we note that all four
algorithms are similarly efficient with a slight advantage for the
GMRES$_\text{ap}$.

Let us conclude this section with the remark that a comparison of the above
algorithms apparently depends very much on the detailed situation in which the
algorithms are used and the specific viewpoint one takes. For example, the
conclusion will be different for the reasons discussed above depending on
whether a simulation is done on a large or intermediate lattice volume, or
whether one is interested in small or intermediate bare quark masses. In a
quenched or partially quenched calculation one will be interested in MM
algorithms which e.g.~would exclude the GMRES$_\text{ap}$ and MR$_\text{ap}$,
on the other hand in a dynamical simulation this exclusion is only important
when a RHMC algorithm is used \cite{Clark:2005sq,Clark:2006fx}.

\clearpage

%%%%%%%%%%%%%%%%%%%%%%%%%%%%%%%%%%%%%%%%%%%%%%%%%%%%%%%%%%%%%%%%%%%%%%
\subsubsection{Low mode preconditioning}
\label{subsubsec:lmp}
%%%%%%%%%%%%%%%%%%%%%%%%%%%%%%%%%%%%%%%%%%%%%%%%%%%%%%%%%%%%%%%%%%%%%%
Let us now turn to low mode preconditioning. We concentrate on the
non-hermitian LMP versions of GMRES$_\text{ap}$ and MR$_\text{ap}$
(cf.~sec.\ref{subsec:lmp overlap}) and compare it to the hermitian LMP version
of CG$_\text{ap}$ \cite{Giusti:2002sm} and in the following we denote the LMP
version of the algorithms by the additional subscript $_\text{lmp}$. Both
GMRES$_\text{ap,lmp}$ and MR$_\text{ap,lmp}$ are particularly promising since
the un-preconditioned versions show a rather bad performance towards small
quark masses, i.e.~they fail to perform efficiently if the condition number of
the operator $D$ gets too large. Obviously, projecting out the few lowest
modes of $D$ and treating them exactly essentially keeps the condition number
constant even when the bare quark mass is pushed to smaller values, e.g.~into
the $\varepsilon$--regime, and hence it has the potential to be particularly
useful. Furthermore, we expect the iteration numbers to decrease for the LMP
operators so that the overhead of GMRES and MR with respect to CG due to the
way the adaptive precision is implemented becomes less severe.

The low modes are calculated using the methods described in appendix
\ref{app:ev}. For the following comparison the normalised low modes
$\psi^{(\pm)}_k$ of $A_\pm = P_\pm D^\dagger D P_\pm$ are calculated
separately in each chiral sector up to a precision which is defined through
the norm of the gradient $g^{(\pm)}_k$ in analogy to Eq.(\ref{eq:right
  gradient}). For later convenience we introduce the triplet notation
$(n_0,n_+,n_-)$ to indicate the set of $n_0$ zero modes and $n_\pm$ modes in
the chirally positive and negative sector, respectively.  These eigenvectors
can directly be used in the CG$_\text{ap,lmp}$, but for the
GMRES$_\text{ap,lmp}$ and MR$_\text{ap,lmp}$ one has to reconstruct the
approximate (left and right) eigenvectors, eigenvalues and gradients. This is
achieved by diagonalising the operator $D$ in the subspace spanned by the
modes $\psi^{(\pm)}_k$ leading to Eq.(\ref{eq:right gradient}) and
(\ref{eq:left gradient}).

\begin{figure}[htb]
  \begin{center}
    \includegraphics[width=10.0cm]{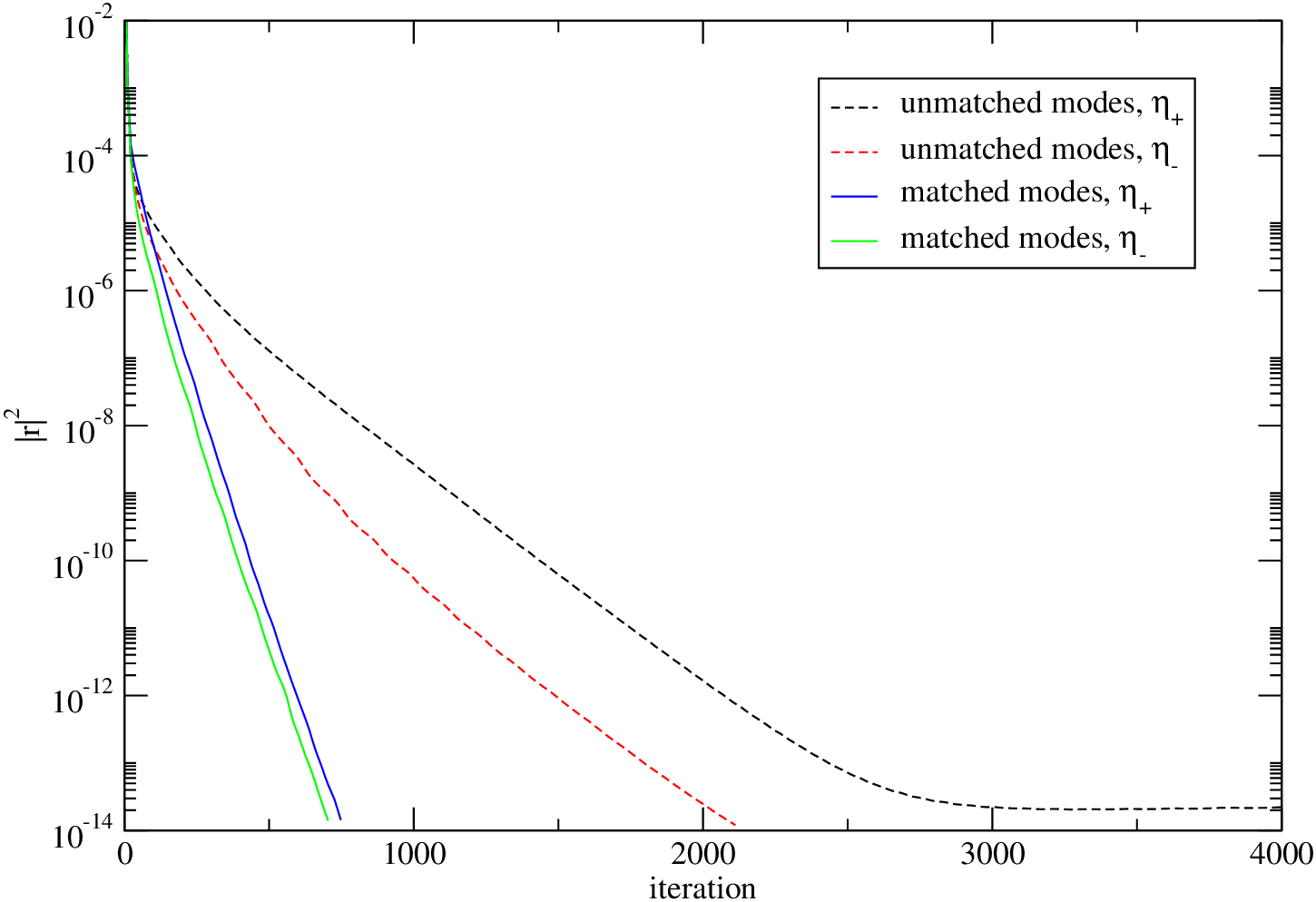}
  \end{center}
  \caption{The square norm of the true
    residual $|r|^2$ of the LM preconditioned operator against the iteration
    number of the GMRES$_\text{ap,lmp}(10)$ algorithm at $\mu=0.005$ on a
    sample $16^4$ configuration with topological index $\nu=5$. The full lines
    show the convergence when the modes of the two chiral sectors are matched,
    $n_+=n_-$, dashed lines when the modes are not matched, $n_+ \neq n_-$.
    $\eta_\pm$ refers to the chirality of the point source.}
  \label{fig:gmres_matched-non-matched}
\end{figure}

At this point it appears to be important that the number of modes $n_\pm$ in
the two chiral sectors match each other (up to zero modes of the massless
operator) in order for the non-hermitian LMP to work efficiently. This is
illustrated in figure \ref{fig:gmres_matched-non-matched} where we plot the
square norm of the true residual $|r|^2$ of the preconditioned operator
Eq.(\ref{eq:A_lmp}) against the iteration number of the
GMRES$_\text{ap,lmp}(10)$ algorithm at $\mu=0.005$ on a sample $16^4$
configuration with topological index $\nu=5$.  The two full lines show the
residuals in the case where the set $(5,10,10)$ is used while the dashed lines
are the residuals obtained with the set (5,5,12). So in addition to the five
zero modes, in the latter case only the first five non-zero modes of the
non-hermitian operator can effectively be reconstructed while in the former
case it is the first 10 non-zero modes leading to a much improved convergence.
More severe, however, is the fact that the convergence may become unstable if
the modes are not matched.

\begin{figure}[htb]
\begin{center}
\includegraphics[width=10.0cm]{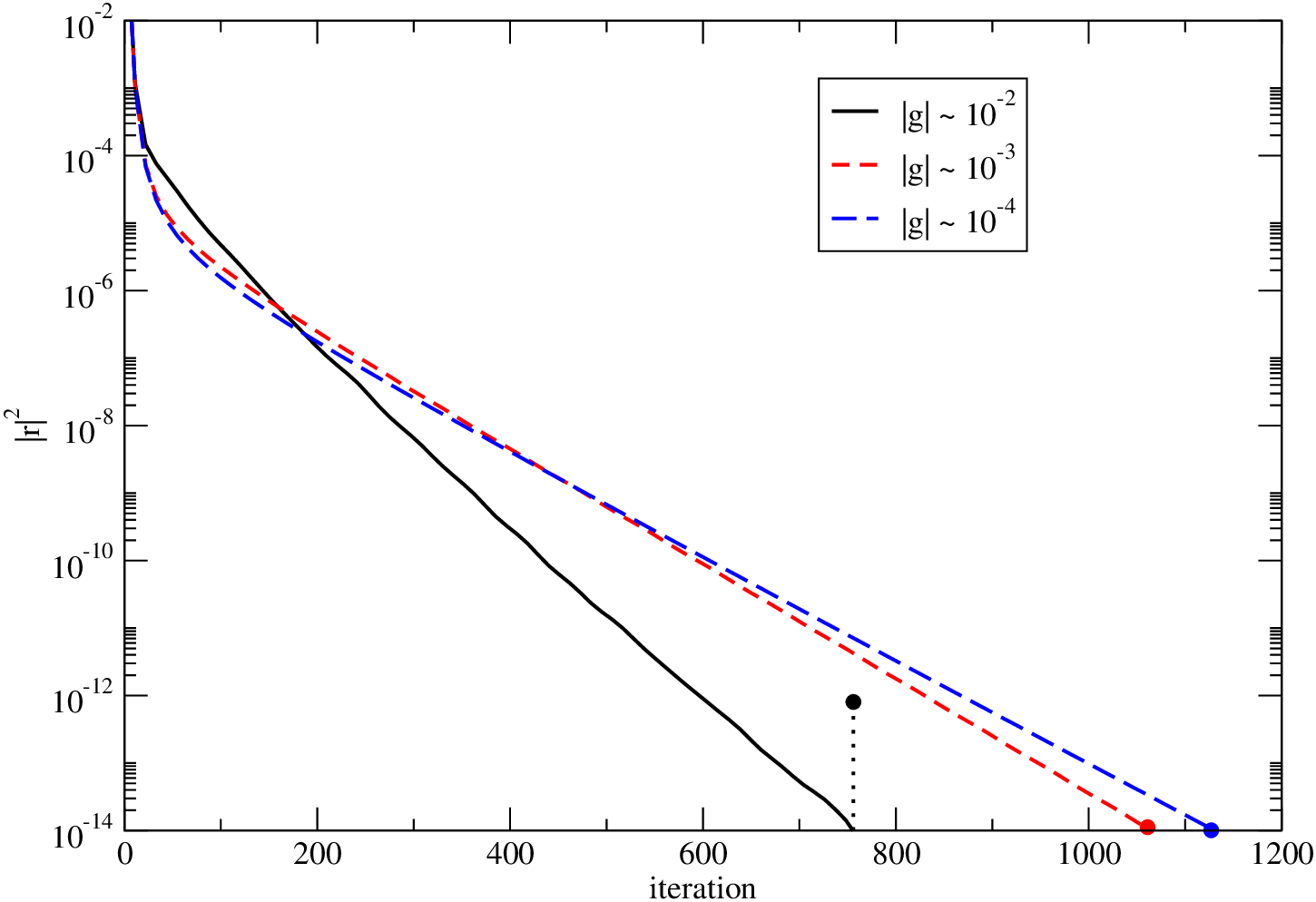}
\end{center}
\caption{The square norm of the true
  residual $|r|^2$ of the LM preconditioned operator against the iteration
  number of the GMRES$_\text{ap,lmp}(10)$ algorithm at $\mu=0.005$ on a sample
  $16^4$ configuration with topological index $\nu=5$. For the low mode set
  (5,10,10) calculated to an accuracy of $|g^{(\pm)}_k|^2 \simeq 10^{-4}$
  (solid black), $10^{-6}$ (short dashed red) and $10^{-8}$ (long dashed
  blue). The dot denotes the residual after adding in the contribution from
  the low modes.  }\label{fig:lmp_gmres_conv_m0.005_1e-2_1e-3_1e-4}
\end{figure}

In the example above we have used modes $\psi^{(\pm)}_k$ that were calculated
with an accuracy $|g^{(\pm)}_k|^2 \lesssim 10^{-4}$ which, after the
reconstruction of the $l_k$ and $r_k$'s, translated into $|g^{(l,r)}_k|^2
\simeq 5 \cdot 10^{-3}$. It is surprising to see that the LMP works even with
such a low accuracy of the low modes. On the other hand, after convergence of
the LM preconditioned operator (cf.~eq.(\ref{eq:A_lmp})) and after adding in
the contributions from the low modes (cf.~eq.(\ref{eq:P psi})), we find that
there is a loss in the true residual of the full operator. This is illustrated
in figure \ref{fig:lmp_gmres_conv_m0.005_1e-2_1e-3_1e-4}
where we show the square norm of the true residual $|r|^2$ of the LM
preconditioned operator against the iteration number of the
GMRES$_\text{ap,lmp}(10)$ algorithm for the same configuration as before,
using the set (5,10,10) calculated to an accuracy of $|g^{(\pm)}_k| \simeq
10^{-4}$ (solid black line) together with the true residual (filled black
circle) after adding in the contribution from the low modes.  On the other
hand, if we use the set (5,10,10) calculated to an accuracy of $|g^{(\pm)}_k|
\simeq 10^{-6}$ (short dashed red line) and $10^{-8}$ (long dashed blue line),
the true residual can be sustained at $|r|^2 \simeq 10^{-14}$ even after
adding in the low mode contributions (filled circles). What is surprising,
however, is that the version using the least accurate low modes converges the
fastest, while the version using the most accurate low modes converges
slowest.

\begin{figure}[htb]
\begin{center}
\includegraphics[width=10.0cm]{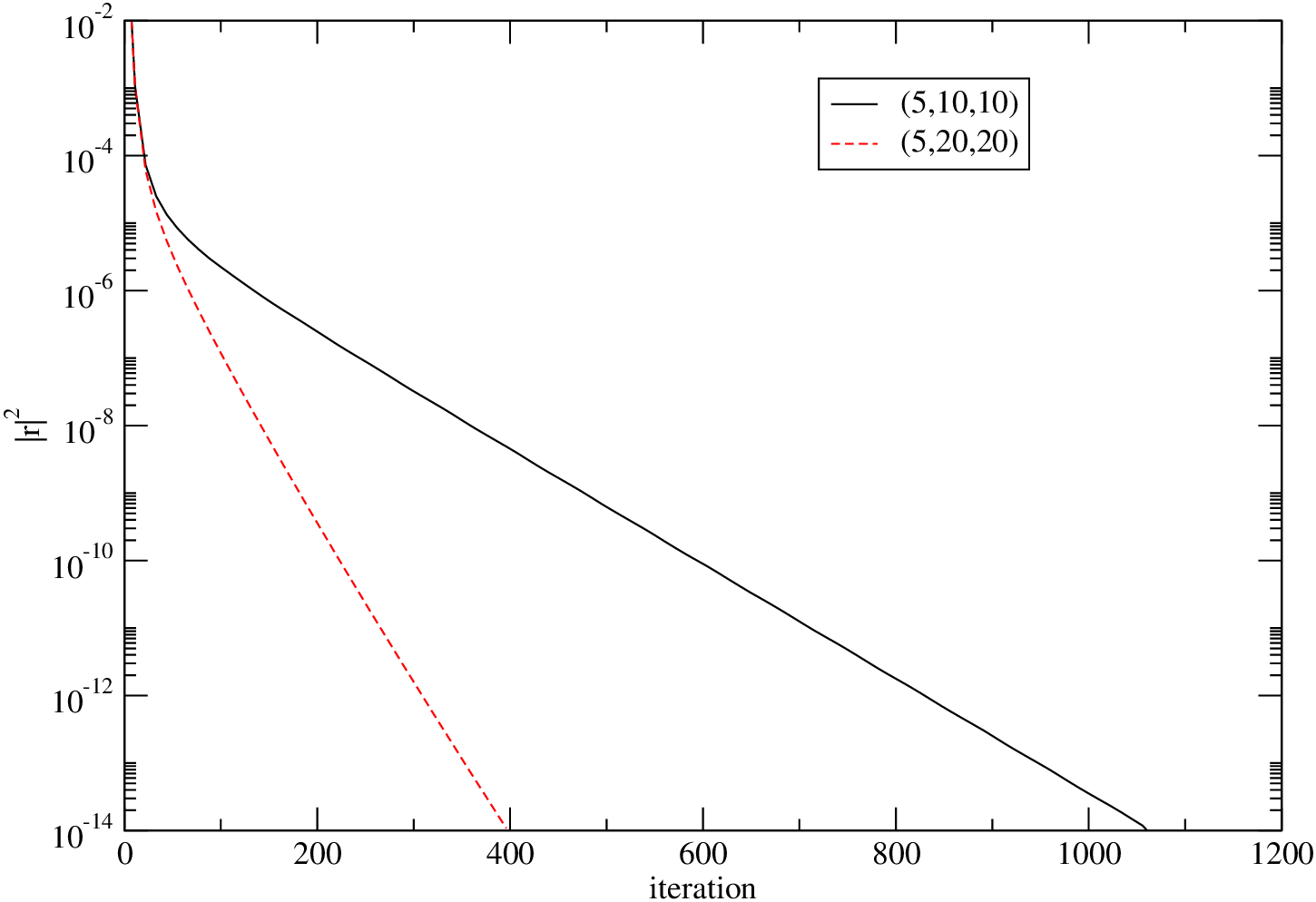}
\end{center}
\caption{The square norm of the true
  residual $|r|^2$ of the LM preconditioned operator against the iteration
  number of the GMRES$_\text{ap,lmp}(10)$ algorithm at $\mu=0.005$ on a sample
  $16^4$ configuration with topological index $\nu=5$. For the low mode set
  (5,10,10) (solid black line) and (5,20,20) (short dashed red line) matched
  low modes calculated to an accuracy of $|g^{(\pm)}_k|^2 \simeq 10^{-6}$.
}\label{fig:lmp_gmres_conv_m0.005_1e-3_10_20}
\end{figure}

Another point worth investigating is how the convergence depends on the number
of projected modes. In figure \ref{fig:lmp_gmres_conv_m0.005_1e-3_10_20} we
show the convergence history for
the GMRES$_\text{ap,lmp}(10)$ algorithm in the case when the set (5,10,10)
(solid black line) and (5,20,20) (short dashed red line) of low modes
calculated to an accuracy of $|g^{(\pm)}_k|^2 \simeq 10^{-6}$ are used for the
preconditioning. In both cases the convergence is approximately exponential
with exponents 0.0195 and 0.056 for the preconditioning with (5,10,10) and
(5,20,20) modes, respectively, and the ratio of exponents matches precisely
the ratio of the squared condition numbers of the preconditioned operators.
Finally we note that there is no sensitivity to whether or not the source is
in the chiral sector which contains the zero-modes.

\begin{figure}[htb]
\begin{center}
\includegraphics[width=10.0cm]{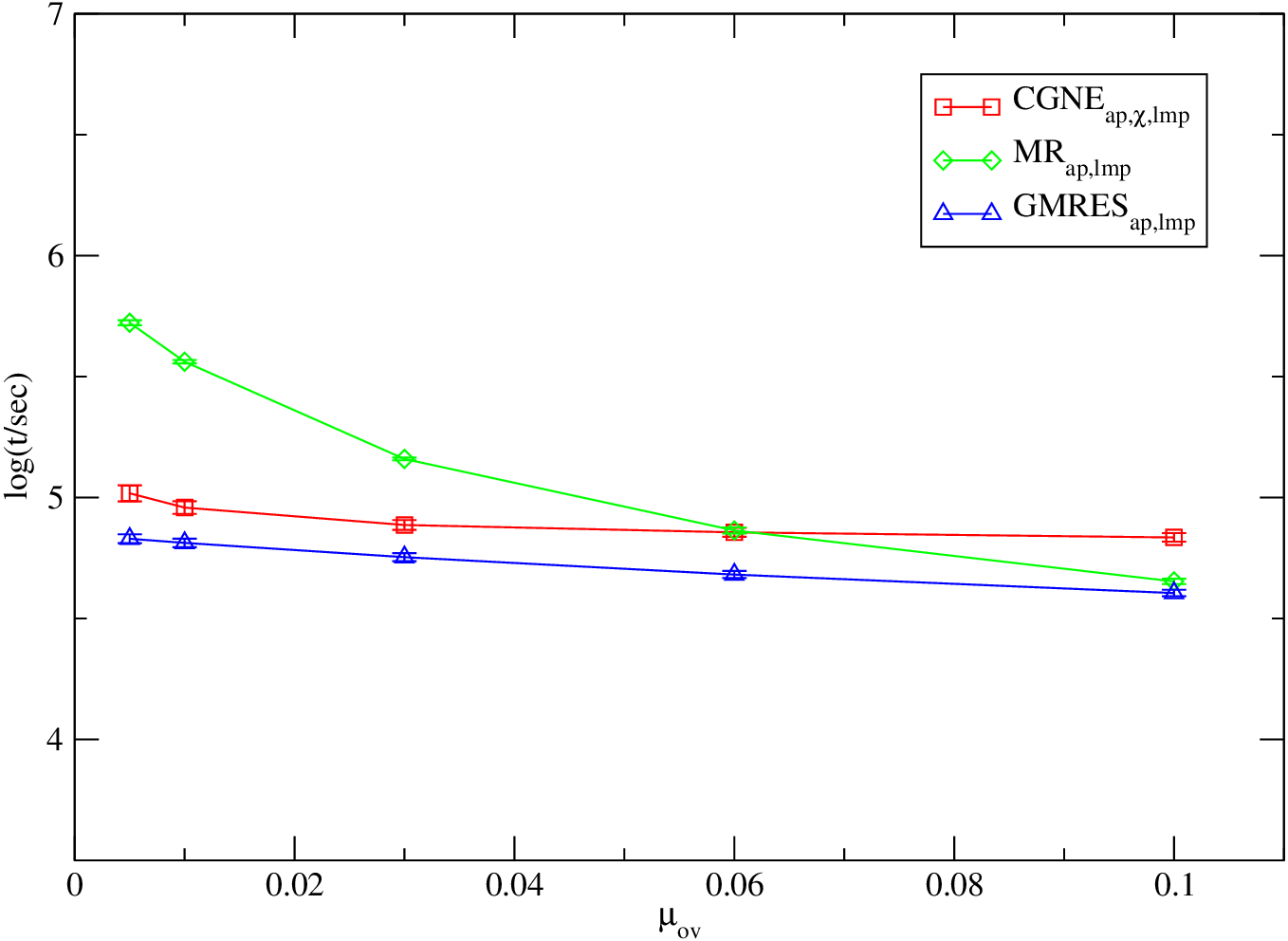}\\
\includegraphics[width=10.0cm]{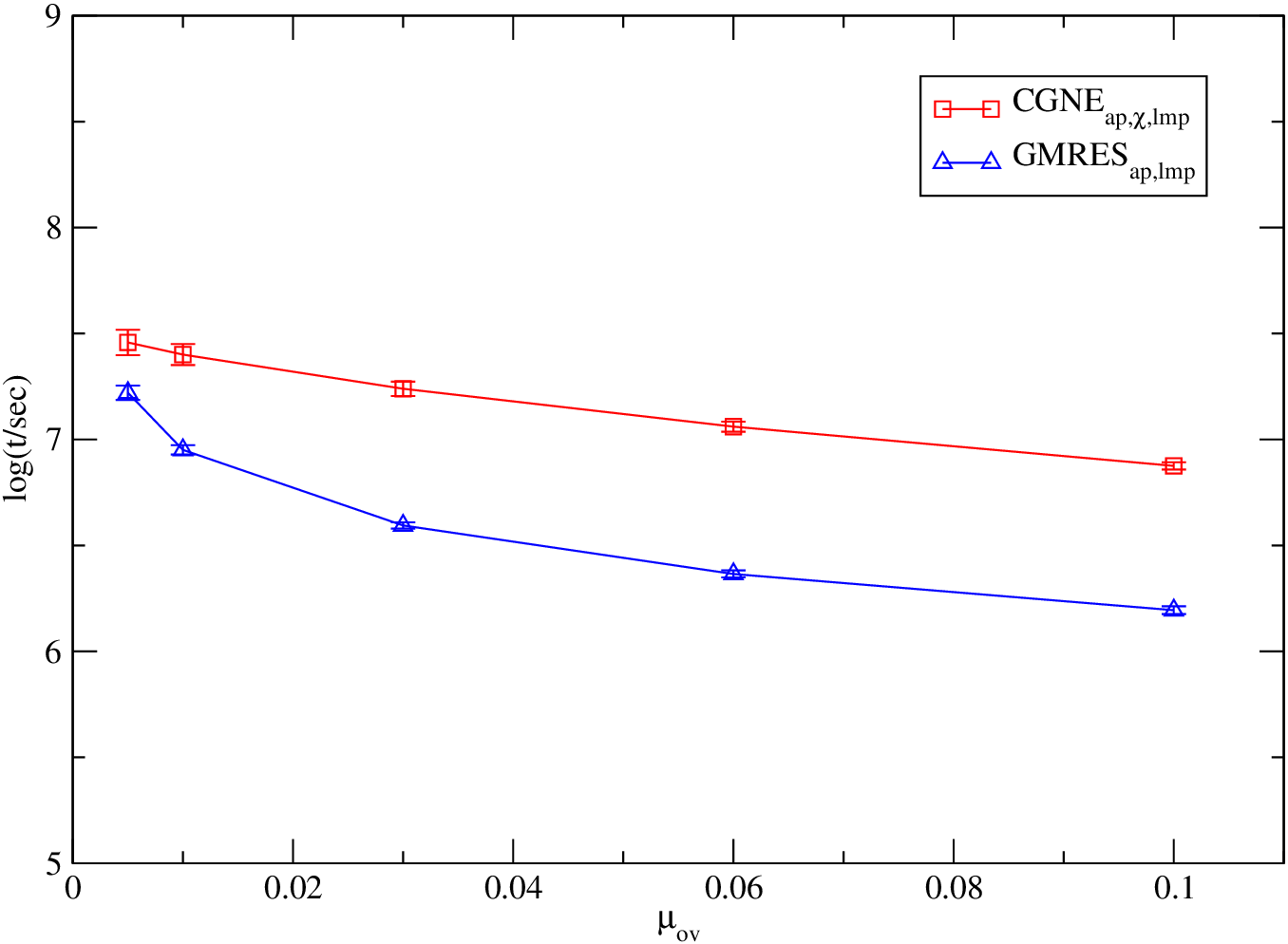}
\end{center}
\caption{Average timings for the inversion of the low mode preconditioned 
  overlap operator in units of seconds for different bare quark masses. At
  $\beta=5.85$ on the $12^4$ (top) and the $16^4$ lattice (bottom).}
\label{fig:timings_lmp_overlap}
\end{figure}

In fig.\ref{fig:timings_lmp_overlap} we show the average timings for the
inversion of the LM preconditioned overlap operator.  For the LMP in addition
to the zero modes we have used 10 nonzero modes on both volumes, i.e.~the set
$(n_0,10,10)$. Obviously, to achieve similar improvement on different volumes
one should scale the number of low modes with the volume. The fact that we
have not done so is reflected in the degradation of the algorithm performance
on the larger volume towards smaller quark mass, but one should keep in mind
that the improvement w.r.t.~the un-preconditioned operator can be easily
enhanced by using more low modes.

The scale is chosen so that the figures can be directly compared to the ones
in fig.\ref{fig:log_timings_overlap}, but we remark that such a comparison is
only of limited interest, since the improvement w.r.t.~the un-preconditioned
operator will depend strongly on the number of low modes and the quark mass.

The timings include all the preparation of the eigenmodes as described in
section \ref{subsubsec:lmp}. Comparing the results for the highest mass with
the ones in fig.\ref{fig:log_timings_overlap} it becomes clear that the
preparation amounts to a non-negligible fraction of the total time, but it
should be noted that in a real production it has to be done only once for all
inversions on a given configuration.

In conclusion we find that GMRES$_\text{ap,lmp}$ outperforms
CG$_\text{ap,lmp}$ by factors of up to two in the range of parameters
investigated here. Due to the favourable volume scaling of the
GMRES$_\text{ap,lmp}$ algorithm this factor is expected to become even larger
on larger volumes.

\clearpage

%%%%%%%%%%%%%%%%%%%%%%%%%%%%%%%%%%%%%%%%%%%%%%%%%%%%%%%%%%%%%%%%%%%%%%
\subsection{Comparison between overlap and Wilson TM}
\label{sec:Comparison overlap TM}
%%%%%%%%%%%%%%%%%%%%%%%%%%%%%%%%%%%%%%%%%%%%%%%%%%%%%%%%%%%%%%%%%%%%%%
The results of the previous sections emphasise that an investigation like the
present one is worthwhile -- for both the overlap and the twisted mass
operator the relative cost factor between the worst and the best algorithm can
be as much as one order of magnitude.

Let us compare directly the absolute and relative cost for the overlap and
twisted mass operator in table \ref{tab:comp number of Q applications} and
table \ref{tab:comp seconds} where we pick in each case the best available
algorithm, GMRES$_\text{ap}$ for the overlap and CGS$_\text{eo}$ for the
twisted mass operator.  We observe that the relative factor in the cost
(measured in execution time or MV products) lies between 30 for the heaviest
mass under investigation and 120 for the lightest mass. The pattern appears to
be similar for the two volumes we looked at, even though the relative factor
is slightly increasing with the volume.

\begin{table}[thb]
  \begin{center}
    \begin{tabular*}{.8\textwidth}{@{\extracolsep{\fill}}rccc}
      \hline\hline
      $V,m_\pi$[MeV] &  Wilson TM  & overlap & rel. factor \\ 
      \hline
      $12^4, 720$ & 599  & 18.1  &  30 \\
      555         & 774  & 27.9  &  36 \\
      390         & 944  & 52.9  &  56 \\
      230         & 1048 & 96.7  &  92 \\
      \hline                                   
      $16^4, 720$ & 650  & 22.3  &  34 \\
      555         & 962  & 34.5  &  36 \\
      390         & 1317 & 66.0  &  50 \\
      230         & 1687 & 198.5 &  118 \\
      \hline\hline        
    \end{tabular*}
    \caption{
      Number of $Q$ applications for the best available algorithm and the
      corresponding relative cost factor. For the overlap operator the number
      of $Q$ applications is in units of 1000.}
    \label{tab:comp number of Q applications}
  \end{center}
\end{table}

\begin{table}[thb]
  \begin{center}
    \begin{tabular*}{.8\textwidth}{@{\extracolsep{\fill}}rccc}
      \hline\hline
      $V,m_\pi$[MeV] &  Wilson TM  & overlap & rel. factor \\ 
      \hline
      $12^4, 720$ &    1.0  &     48.8 &  49 \\
      555         &    1.3  &     75.1 &  58 \\
      390         &    1.6  &    141.5 &  88 \\
      230         &    1.8  &    225.0 & 125 \\
      \hline                                         
      $16^4, 720$ &    3.7  &    225.3 &  61 \\
      555         &    5.2  &    343.9 &  66 \\
      390         &    6.8  &    652.7 &  96 \\
      230         &   10.0  &   1949.3 & 195 \\
      \hline\hline    
    \end{tabular*}
    \caption{
      Absolute timings in seconds on one node of JUelich
      MultiProzessor (JUMP) IBM p690 Regatta in J{\"u}lich for the best
      available algorithm and the corresponding relative cost factor.}
    \label{tab:comp seconds}
  \end{center}
\end{table}

We would like to emphasise that the overlap operator as used in this paper
obeys lattice chiral symmetry up to machine precision and hence the relative
factor compared to TM fermions will be less if a less stringent
Ginsparg-Wilson fermion is used. Including those fermions as well as improved
overlap fermions (for instance with a smeared kernel) in the tests are,
however, beyond the scope of this paper.

%%%%%%%%%%%%%%%%%%%%%%%%%%%%%%%%%%%%%%%%%%%%%%%%%%%%%%%%%%%%%%%%%%%%%%
%%%%%%%%%%%%%%%%%%%%%%%%%%%%%%%%%%%%%%%%%%%%%%%%%%%%%%%%%%%%%%%%%%%%%%
\section{Conclusions and Outlook} 
\label{sec:Conclusions and Outlook}
%%%%%%%%%%%%%%%%%%%%%%%%%%%%%%%%%%%%%%%%%%%%%%%%%%%%%%%%%%%%%%%%%%%%%%
%%%%%%%%%%%%%%%%%%%%%%%%%%%%%%%%%%%%%%%%%%%%%%%%%%%%%%%%%%%%%%%%%%%%%%
In this paper we have performed a comprehensive, though not complete test of
various algorithms to solve very large sets of linear systems employing sparse
matrices as needed in applications of lattice QCD.  We considered two
relatively new formulations of lattice QCD, chirally improved Wilson twisted
mass fermions at full twist and chirally invariant overlap fermions.  The
tests were performed on $12^4$ and $16^4$ lattices and four values of the
pseudo scalar mass of 230MeV, 390MeV, 555MeV and 720MeV.  The lattice spacing
has been fixed to $a\approx 0.125$fm.

We think that our study will help to select a good linear system solver for
twisted mass and overlap fermions for practical simulations.  We emphasise
that we cannot provide a definite choice of the optimal algorithm for each
case.  The reason simply is that the optimal choice depends on many details of
the problem at hand such as the exact pseudo scalar mass, the volume, the
source vector etc..  Nevertheless, in general we find that for twisted mass
fermions CGS appears to be the fastest linear solver algorithm while for
overlap fermions it is GMRES$_\text{ap}$ for the parameters investigated here.
In a direct competition between twisted mass and overlap fermions the latter
are by a factor of 30-120 more expensive if one compares the best available
algorithms in each case with an increasing factor when the value of the pseudo
scalar mass is lowered. Preconditioning plays an important role for both
investigated fermion simulations. A factor of two is obtained by using
even/odd preconditioning for the TM operator. A similar improvement can be
expected from SSOR preconditioning \cite{Fischer:1996th,Frommer:1994vn}.

For the overlap operator it turns out to be rather efficient to adapt the
precision of the polynomial approximation in the course of the solver
iterations. This easily speeds up the inversion by a factor of two. In the
$\epsilon$-regime in addition low mode preconditioning can overcome the
slowing down of the convergence of the algorithms towards small quark masses
and the convergence rate can essentially be kept constant for all masses. In
particular we find that the GMRES$_\text{ap,lmp}$ outperforms
CG$_\text{ap,lmp}$ by factors of up to two with tendency of getting even
better towards larger volumes.

One of the aims of this paper has been to at least start an algorithm
comparison and we would hope that our study here will be extended by other
groups adding their choice of algorithm, optimally {\em using the here
  employed simulations parameters as benchmark points.}  In this way, a
toolkit of algorithms could be generated and gradually enlarged.

\section*{Acknowledgements}

We thank NIC and the computer centre at Forschungszentrum J{\"u}lich
for providing the necessary technical help and computer resources.
This work has been supported in part by the DFG
Sonderforschungsbereich/Transregio SFB/TR9-03 and the EU Integrated
Infrastructure Initiative Hadron Physics (I3HP) under contract
RII3-CT-2004-506078. We also thank the DEISA Consortium (co-funded by
the EU, FP6 project 508830), for support within the DEISA Extreme
Computing Initiative (www.deisa.org). The work of T.C. is supported
by the DFG in the form of a Forschungsstipendium CH 398/1.

%%%%%%%%%%%%%%%%%%%%%%%%%%%%%%%%%%%%%%%%%%%%%%%%%%%%%%%%%%%%%%%%%%%%%%
%%%%%%%%%%%%%%%%%%%%%%%%%%%%%%%%%%%%%%%%%%%%%%%%%%%%%%%%%%%%%%%%%%%%%%
\begin{appendix}
%%%%%%%%%%%%%%%%%%%%%%%%%%%%%%%%%%%%%%%%%%%%%%%%%%%%%%%%%%%%%%%%%%%%%%
\section{Eigenpair Computation}
\label{app:ev}
%%%%%%%%%%%%%%%%%%%%%%%%%%%%%%%%%%%%%%%%%%%%%%%%%%%%%%%%%%%%%%%%%%%%%%
%%%%%%%%%%%%%%%%%%%%%%%%%%%%%%%%%%%%%%%%%%%%%%%%%%%%%%%%%%%%%%%%%%%%%%
As mentioned already in section~\ref{sec:Lattice Dirac operators}, the
computation of eigenvalues and eigenvectors or approximations of those are
needed in various methods used in this paper, e.g.~for the practical
implementation of the $\sign$-function or the low mode preconditioning of the
overlap operator. But also if one is interested in computing the topological
index with the overlap operator one needs an algorithm to compute the
eigenvalues of the overlap operator.

The standard method used in lattice QCD is the so called Ritz-Jacobi method
\cite{Kalkreuter:1995mm}. For the use of adaptive precision for the overlap
operator with this method, see Ref.~\cite{Cundy:2002hv,Giusti:2002sm}.
Another choice would be the Arnoldi algorithm implemented in the
ARPACK package which, however, sometimes fails to compute for instance
a given number 
of the lowest eigenvalues of $Q^2$ by missing one. This might lead to problems
if the eigenvalues are used to normalise the Wilson-Dirac operator in the
polynomial construction of the overlap operator.

We used yet another method which is described in the following section. After
that we present some implementation details for the determination of the
index.

%%%%%%%%%%%%%%%%%%%%%%%%%%%%%%%%%%%%%%%%%%%%%%%%%%%%%%%%%%%%%%%%%%%%%%
\subsection{Jacobi-Davidson method}
%%%%%%%%%%%%%%%%%%%%%%%%%%%%%%%%%%%%%%%%%%%%%%%%%%%%%%%%%%%%%%%%%%%%%%
Consider a complex valued $N\times N$ matrix $A$ for which we aim at
determining (part of) its eigenvalues and eigenvectors. The exact computation
of those is in general too demanding and thus one has to rely on some
iterative method. The one we are going to describe here was introduced in
\cite{Sleijpen:1996aa}.

Assume we have an approximation $(\lambda_k, u_k)$ for the eigenpair
$(\lambda, u)$ and we want to find a correction $v$ to $u_k$ in order to
improve the approximation. One way of doing this is to look for the orthogonal
complement for $u_k$ with respect to $u$, which means we are interested in the
subspace $u_k^\perp$.

The projection of $A$ into this subspace is given by
\begin{equation}
  \label{jd:0}
  B_k \equiv (I-u_k u_k^\dagger)A(I-u_k u_k^\dagger)\, ,
\end{equation}
where the vector $u_k$ has been normalised and $I$ represents the
identity matrix. Eq.~(\ref{jd:0}) can be rewritten as follows
\begin{equation}
  \label{jd:1}
  A = B_k + A u_k u_k^\dagger + u_k u_k^\dagger A - \lambda_k u_k
  u_k^\dagger\, .
\end{equation}
Since we want to find $v \perp u_k$ such that
\[
A(u_k + v) = \lambda(u_k + v)\, ,
\]
it follows with $B_k u_k=0$
\begin{equation}
  \label{jd:2}
  (B_k-\lambda I)v = -r_k + (\lambda -\lambda_k - u_k^\dagger A v)u_k\,
  ,
\end{equation}
where we introduced the residual vector $r_k$ given by
\[
r_k = (A - \lambda_k I)u_k\, .
\]
Neither $r_k$ nor the l.h.s of Eq.~(\ref{jd:2}) have a component in
direction $u_k$ and hence $v$ should satisfy
\begin{equation}
  \label{jd:3}
  (B_k - \lambda I) v = -r_k\, .
\end{equation}
Since $\lambda$ is unknown, we replace it by $\lambda_k$ and Eq.~(\ref{jd:3})
can then be solved with any iterative solver. Note that the matrix $B$ depends
on the approximation $u_k$ and needs to be newly constructed in every step.

Solving Eq.~(\ref{jd:3}) for $v$ in every iteration step might look as if the
proposed algorithm is rather computer time demanding. But it turns out that in
fact it has to be solved only approximately, i.e. in each iteration step only a
few iterations of the solver have to be performed.

\begin{algorithm}
  \caption{Basic Jacobi-Davidson algorithm}
  \label{alg1}
  \begin{algorithmic}[1]
    \REQUIRE non trivial initial guess vector $v$, $m>1$
    \vspace{.2cm}
    \STATE $v_1 = v/\|v\|$, $w_1 = Av_1$, $h_{11} = v_1^\dagger w_1$, $i=1$
    \STATE $V_1 = \{v_1\}$, $W_1 = \{w_1\}$, $H_1 = \{h_{11}\}$
    \STATE $u_k = v_1$, $\lambda_1 = h_{11}$
    \STATE $r_1 = w_1 - \lambda_1 u$
    \vspace{.2cm}
    \REPEAT
    \FOR{$k=1,...,m-1$}
    \STATE solve approximately for $v_{k+1}\perp u_k$
    \begin{equation}
      \label{jd:4}
      (I-u_k u_k^\dagger)(A-\lambda_k I)(I-u_k u_k^\dagger)v_{k+1} = -r_k
    \end{equation}
    \STATE orthogonalise $v_{k+1}$ against $V_k$, $V_{k+1} = \{V_k, v_{k+1}\}$
    \STATE $w_{k+1}=A v_{k+1}$, $W_{k+1} = \{W_k, w_{k+1}\}$
    \STATE compute $V_{k+1}^\dagger w_{k+1}$, the last column of
    $H_{k+1}\equiv V_{k+1}^\dagger A V_{k+1}$
    \IF{$A$ is not hermitian}
    \STATE compute $v_{k+1}^\dagger W_{k}$, the last row of
    $H_{k+1}$
    \ENDIF
    \STATE compute the smallest eigenpair $(\lambda_{k+1},s)$ of
    $H_{k+1}$ and normalise $s$.
    \STATE $u_{k+1} = V_{k+1}s\quad$ \COMMENT{The new eigenvector approximation}
    \STATE $\hat u = A u_{k+1}$ and $r_{k+1} = \hat u -
    \lambda_{k+1} u_{k+1}$
    \STATE test for convergence, $i=i+1$
    \ENDFOR
    \STATE restart: Set $V_1 = \{u_m\}$, $W_1 = \{\hat u\}$, $H_1 = \{\lambda_m\}$
    \UNTIL{convergence}
  \end{algorithmic}
\end{algorithm}

The basic Jacobi-Davidson (JD) algorithm is summarised in algorithm
\ref{alg1}. In algorithm \ref{alg1} we denote matrices with capital letters
and vectors with small letters. $V=\{v\}$ means that the matrix $V$ contains
only one column $v$, while $W=\{V,v\}$ means that $V$ is expanded by $v$ to
the matrix $W$ by one column.  The basic algorithm can be easily extended in
order to compute more than the minimal (maximal) eigenvalue: the simplest way
is to perform a restart and restrict the eigenvector search to the subspace
orthogonal to the already computed eigenvector(s).

A further way to compute more than one eigenvalue is to solve Eq.~(\ref{jd:3})
more than once per iteration for several approximate eigenvectors. This so
called blocking method is also capable to deal with degenerate eigenvalues,
which are otherwise not correctly computed by the JD method
\cite{Fokkema:1998aa,Geus:2002}.

Moreover, the JD algorithm is able to compute eigenpairs which are located in
the bulk eigenvalue spectrum of $A$. This is achieved by replacing $\lambda_k$
in Eq.~(\ref{jd:3}) by an initial guess $\sigma$ in the first few iterations,
which will drive the JD algorithm to compute preferably eigenvalues close to
$\sigma$ \cite{Sleijpen:1996aa}.

Let us finally discuss some implementation details regarding the
parallelisation of the JD algorithm. As soon as the application of $A$ is
parallelised most of the remaining linear algebra operations are parallelised
trivially, which includes matrix-vector multiplications as $V_{k+1}^\dagger
w_{k+1}$. Only the computation of the eigenvalues of the (low dimensional)
matrix $H$ is not immediately parallelisable and, in fact, it is most
efficient to hold a local copy of $H$ and compute the eigenvalues on each
processor. Then the multiplication $V_{k+1} s$ in line $15$ of algorithm
\ref{alg1} is even a local operation. This seems to be a doubling of work, but
as $H$ is only an order $20\times20$ matrix, the parallelisation overhead
would be too large.

%%%%%%%%%%%%%%%%%%%%%%%%%%%%%%%%%%%%%%%%%%%%%%%%%%%%%%%%%%%%%%%%%%%%%%
\subsubsection{Index computation}
%%%%%%%%%%%%%%%%%%%%%%%%%%%%%%%%%%%%%%%%%%%%%%%%%%%%%%%%%%%%%%%%%%%%%%
The computation of the topological index on a given gauge configuration with
the overlap operator involves counting the zero modes of $D_\mathrm{ov}$. More
precisely, the chiral sector containing zero modes has to be identified and
then their number has to be determined. To this end we have implemented the
method of Ref.~\cite{Giusti:2002sm} which makes use of the Ritz-Jacobi
algorithm. Moreover, it is straightforward to adapt the method also to the JD
algorithm. So we are not going to mention the details of this algorithm.

But it is useful to discuss some performance improvements of the index
computation: the most time consuming part in the JD algorithm is to find an
approximate solution to Eq.~(\ref{jd:3}).  As suggested in
Refs.~\cite{Fokkema:1998aa,Geus:2002}, we used the following set-up. The
actual absolute precision to which the solution is driven is computed as
\[
\epsilon = x^{-i}\, ,
\]
where $i$ counts the number of JD-iterations performed so far for the
eigenvalue in question (see algorithm \ref{alg1}) and $x=1.5$. Additionally we
set the maximal number of iterations in the solver to $50$. In this way we
avoid on the one hand that the solution to Eq.~(\ref{jd:3}) is much more
precise than the current approximation for the eigenvalue and on the other
hand too many iterations in the solver.

Thus, most of the time, the precision required from the iterative solver is
only rough, and hence it is useful to use adaptive precision for the
$\sign$-function, since the polynomial approximation of the $\sign$-function
is not needed to be much more precise than the required solver precision. Our
experience shows that setting the precision in the polynomial to
$10^{-2}\cdot\epsilon$ is a good choice in this respect. We remark that the
vector $w_{k+1}$ (see line $9$ in algorithm \ref{alg1}) as well as the next
residual $r_{k+1}$ should be computed with full precision in the polynomial in
order not to bias the computation.

%%%%%%%%%%%%%%%%%%%%%%%%%%%%%%%%%%%%%%%%%%%%%%%%%%%%%%%%%%%%%%%%%%%%%%
\subsection{Index from the CG search}
\label{subsec:indexCG}
%%%%%%%%%%%%%%%%%%%%%%%%%%%%%%%%%%%%%%%%%%%%%%%%%%%%%%%%%%%%%%%%%%%%%%
For the computation of the topological index it is important to note that the
determination of the chiral sector which contains the zero-modes comes for
free when one uses the CG-algorithm for the inversion.  By estimating the
eigenvalues once in each chiral sector and by pairing them accordingly it is
possible to identify the chiral sector which contains zero modes.  From the
CG-coefficients which are obtained during the iteration one can build up a
tridiagonal matrix which is related to the underlying Lanczos procedure
\cite{saad:2003a}. The eigenvalues of this matrix approximate the extremal
eigenvalues of the operator and it turns out that the lowest 5-10 eigenvalues
are approximated rather accurately.
               
\begin{figure}[htb]
  \centering
\includegraphics[angle=0,width=10.0cm]{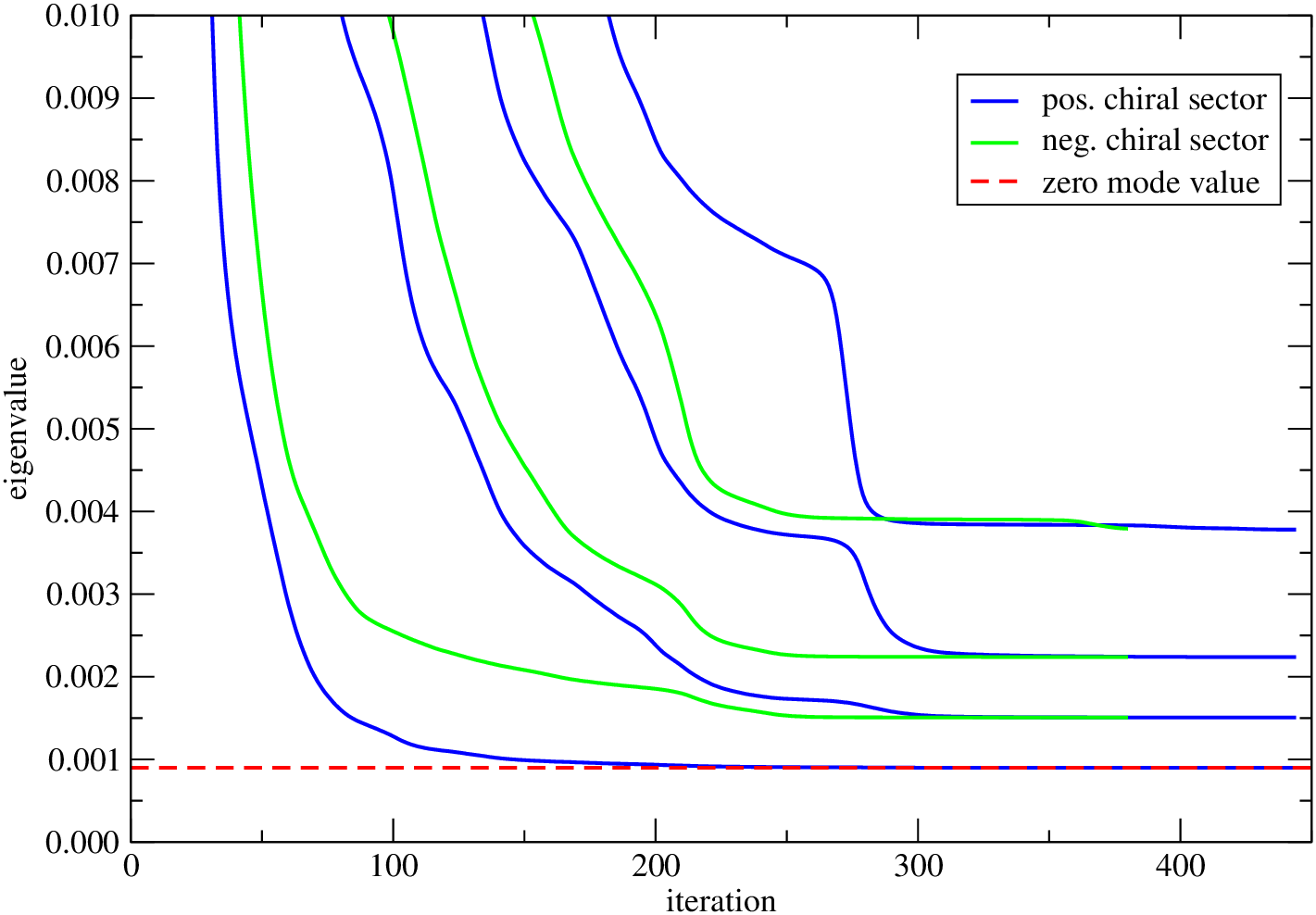}
\includegraphics[angle=0,width=10.0cm]{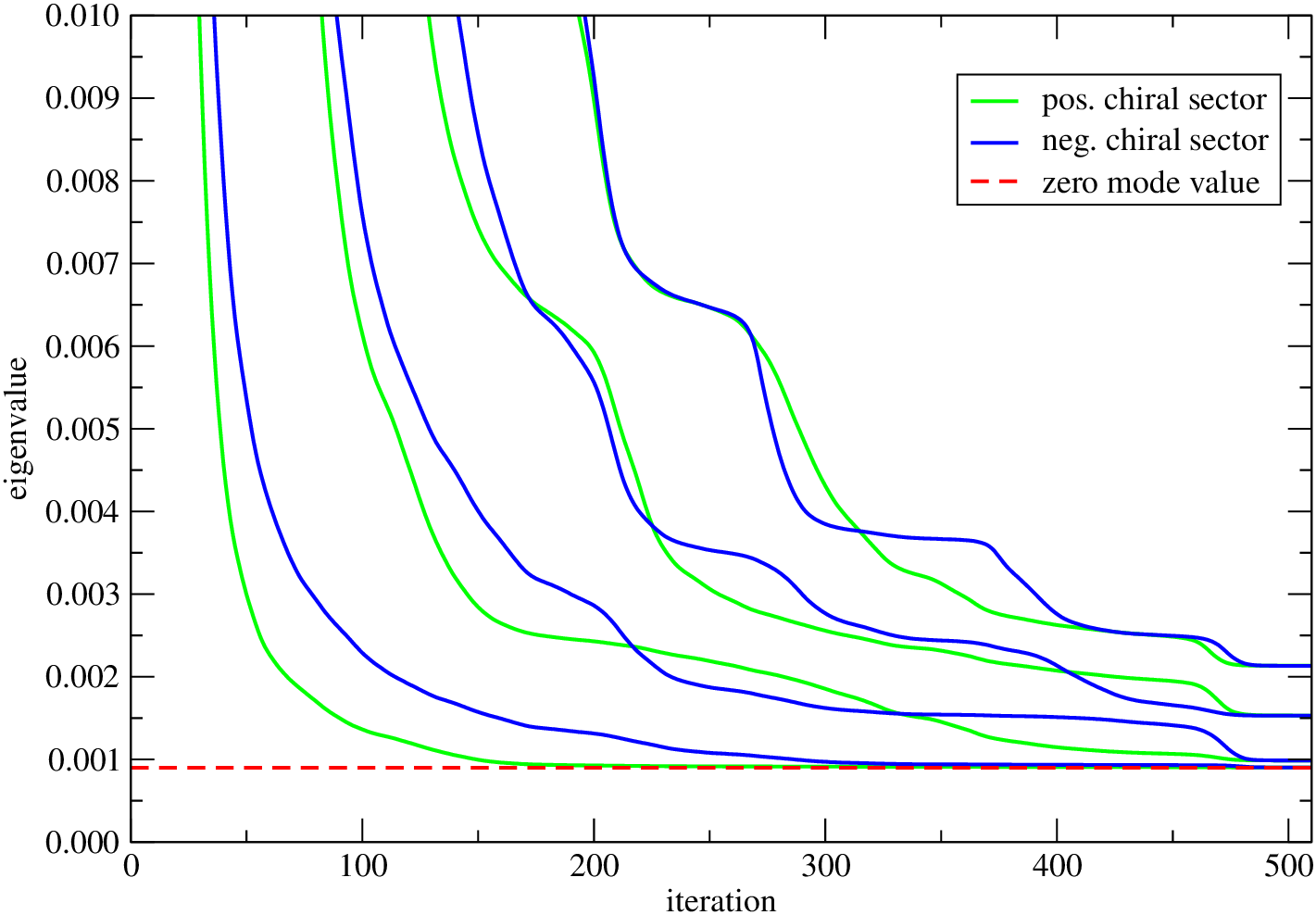}
  \caption{{}Estimates of the lowest eigenvalues of the overlap operator
    from the CG-coefficients for a $16^4$ configuration with topological
    charge $\nu<0$ (top) and $\nu = 0$ (bottom) at $\mu_\textrm{ov}=0.03$.}
  \label{fig:cgev}
\end{figure}

In figure \ref{fig:cgev} we plot the iterative determination of the lowest
eigenvalues during a CG-inversion for two different configurations. For the
first configuration (left plot) we see a rapid convergence of the unpaired
lowest eigenvalue towards the zero mode value suggesting a non-zero
topological charge in the given chiral sector. The second configuration on the
other hand (right plot) also shows a rapid convergence of the lowest mode
towards the zero mode value, but this time it is paired by an equal eigenvalue
in the opposite chiral sector hence suggesting a configuration with zero
topological charge. Figure \ref{fig:cgev} emphasises the point that the
pairing of modes in the two chiral sectors is the crucial ingredient for the
determination of the topological charge sector and not the estimate of the
eigenvalue itself. Indeed, for the second configuration the eigenvalue
estimates converge to a value slightly larger than the zero mode value as one
would expect.

%%%%%%%%%%%%%%%%%%%%%%%%%%%%%%%%%%%%%%%%%%%%%%%%%%%%%%%%%%%%%%%%%%%%%%
\section{Multiple mass solver for twisted mass fermions}
\label{app:MM for TM}
%%%%%%%%%%%%%%%%%%%%%%%%%%%%%%%%%%%%%%%%%%%%%%%%%%%%%%%%%%%%%%%%%%%%%%
We want to invert the TM operator at a certain twisted mass $\mu_0$ obtaining
automatically all the solutions for other twisted masses $\mu_k$ (with
$|\mu_k| \ge |\mu_0|$).  Then, as in Eq.(\ref{eq:Dtm}) the Wilson twisted mass
operator is\footnote{In the following the subscript ${}_{\text{tm}}$
  associated with the bare twisted quark masses $\mu_k$ is suppressed.}
\begin{equation}
D_{\rm tm} = D_\mathrm{W} + i\mu_k\gamma_5\tau^3, \qquad k=1,\ldots,N_m
\end{equation}
where $N_m$ is the number of additional twisted masses.  The operator can be
split as
\begin{equation}
  D_{\rm tm} = D_{\rm tm}^{(0)} + i(\mu_k-\mu_0)\gamma_5\tau^3, \qquad
  D_{\rm tm}^{(0)}  = D_\mathrm{W} + i\mu_0\gamma_5\tau^3 
\end{equation}
The trivial observation is that 
\begin{equation}
D_{\rm tm}D_{\rm tm}^{\dagger} = D_{\rm tm}^{(0)}D_{\rm tm}^{(0)\dagger} + \mu_k^2 - \mu_0^2\;,
\end{equation}
where we have used $\gamma_5D_\mathrm{W}\gamma_5 = D_\mathrm{W}^{\dagger}$.
Now clearly we have a shifted linear system $(A+\sigma_k)x-b=0$ with $A =
D_{\rm tm}^{(0)}D_{\rm tm}^{(0)\dagger}$ and $\sigma_k = \mu_k^2 - \mu_0^2$.
In algorithm \ref{cgm} we describe the CG-M algorithm to solve the problem
$(A+\sigma_k)x-b=0$.  The lower index indicates the iteration steps of the
solver, while the upper index $k$ refers to the shifted problem with
$\sigma_k$. The symbols without upper index refer to mass $\mu_0$.

\begin{algorithm}
  \caption{CG-M algorithm}
  \label{cgm}
  \begin{algorithmic}[1]
    \vspace{.2cm}
    \STATE $n=0, x_0^k = 0, r_0 = p_0 = p_0^k = b$
    \STATE  $\biggl.\biggr.\alpha_{-1} = \zeta_{-1}^k = \zeta_0^k = 1, \beta_0^k = \beta_0 = 0$
    \REPEAT
    \STATE $\alpha_n = (r_n, r_n) / (p_n, A p_n)$
    \STATE $\biggl.\biggr.\zeta_{n+1}^k = (\zeta^k_n  \alpha_{n-1}) / 
      (\alpha_n \beta_n(1 - \zeta_n^k / \zeta^k_{n-1}) + \alpha_{n-1}
      (1-\sigma_k\alpha_n))$
    \STATE $\alpha^k_n = (\alpha_n \zeta_{n+1}^k)/ \zeta_n^k$
    \STATE $\biggl.\biggr.x_{n+1}^k = x_n^k + \alpha_n^k p_n^k$
    \STATE $x_{n+1} = x_n + \alpha_n p_n$
    \STATE $\biggl.\biggr.r_{n+1} = r_n - \alpha_n Ap_n$
    \STATE $\beta_{n+1} = (r_{n+1}, r_{n+1}) / (r_n, r_n)$
    \STATE $\biggl.\biggr.p_{n+1}^k = \zeta_{n+1}^k r_{n+1} + \beta_{n+1}^k p_n^k$
    \STATE $n=n+1$
    \UNTIL{$\|r_n\|<\epsilon$}
  \end{algorithmic}
\end{algorithm}

\end{appendix}

\clearpage

\bibliographystyle{h-physrev4}

\bibliography{imotmf}

\begin{thebibliography}{10}

\bibitem{Luscher:1998pq}
M.~L{\"u}scher,
\newblock Phys. Lett. {\bf B428}, 342 (1998), [hep-lat/9802011].
%%CITATION = HEP-LAT 9802011;%%

\bibitem{Neuberger:1997fp}
H.~Neuberger,
\newblock Phys. Lett. {\bf B417}, 141 (1998), [hep-lat/9707022].
%%CITATION = HEP-LAT 9707022;%%

\bibitem{Neuberger:1998wv}
H.~Neuberger,
\newblock Phys. Lett. {\bf B427}, 353 (1998), [hep-lat/9801031].
%%CITATION = HEP-LAT 9801031;%%

\bibitem{Frezzotti:2003ni}
R.~Frezzotti and G.~C. Rossi,
\newblock JHEP {\bf 08}, 007 (2004), [hep-lat/0306014].
%%CITATION = HEP-LAT 0306014;%%

\bibitem{Bietenholz:2004wv}
\xlf, W.~Bietenholz {\em et~al.},
\newblock JHEP {\bf 12}, 044 (2004), [hep-lat/0411001].
%%CITATION = HEP-LAT 0411001;%%

\bibitem{Ukawa:2002pc}
CP-PACS and JL{QCD}, A.~Ukawa,
\newblock Nucl. Phys. Proc. Suppl. {\bf 106}, 195 (2002).
%%CITATION = NUPHZ,106,195;%%

\bibitem{Jansen:2003nt}
K.~Jansen,
\newblock Nucl. Phys. Proc. Suppl. {\bf 129}, 3 (2004), [hep-lat/0311039].
%%CITATION = HEP-LAT 0311039;%%

\bibitem{Urbach:2005ji}
C.~Urbach, K.~Jansen, A.~Shindler and U.~Wenger,
\newblock Comput. Phys. Commun. {\bf 174}, 87 (2006), [hep-lat/0506011].
%%CITATION = HEP-LAT 0506011;%%

\bibitem{saad:2003a}
Y.~Saad,
\newblock {\em Iterative Methods for sparse linear systems}, 2nd ed. (SIAM,
  2003).

\bibitem{Frommer:1995ik}
A.~Frommer, B.~Nockel, S.~Gusken, T.~Lippert and K.~Schilling,
\newblock Int. J. Mod. Phys. {\bf C6}, 627 (1995), [hep-lat/9504020].
%%CITATION = HEP-LAT 9504020;%%

\bibitem{Chiarappa:2004ry}
T.~Chiarappa {\em et~al.},
\newblock Nucl. Phys. Proc. Suppl. {\bf 140}, 853 (2005), [hep-lat/0409107].
%%CITATION = HEP-LAT 0409107;%%

\bibitem{Arnold:2003sx}
G.~Arnold {\em et~al.},
\newblock hep-lat/0311025.
%%CITATION = HEP-LAT 0311025;%%

\bibitem{Cundy:2004pz}
N.~Cundy {\em et~al.},
\newblock Comput. Phys. Commun. {\bf 165}, 221 (2005), [hep-lat/0405003].
%%CITATION = HEP-LAT 0405003;%%

\bibitem{Krieg:2004xg}
S.~Krieg {\em et~al.},
\newblock Nucl. Phys. Proc. Suppl. {\bf 140}, 856 (2005), [hep-lat/0409030].
%%CITATION = HEP-LAT 0409030;%%

\bibitem{Frezzotti:1999vv}
R.~Frezzotti, P.~A. Grassi, S.~Sint and P.~Weisz,
\newblock Nucl. Phys. Proc. Suppl. {\bf 83}, 941 (2000), [hep-lat/9909003].
%%CITATION = HEP-LAT 9909003;%%

\bibitem{Frezzotti:2000nk}
ALPHA, R.~Frezzotti, P.~A. Grassi, S.~Sint and P.~Weisz,
\newblock JHEP {\bf 08}, 058 (2001), [hep-lat/0101001].
%%CITATION = HEP-LAT 0101001;%%

\bibitem{Jansen:2003ir}
\xlf, K.~Jansen, A.~Shindler, C.~Urbach and I.~Wetzorke,
\newblock Phys. Lett. {\bf B586}, 432 (2004), [hep-lat/0312013].
%%CITATION = HEP-LAT 0312013;%%

\bibitem{Jansen:2005kk}
{\xlf}, K.~Jansen, M.~Papinutto, A.~Shindler, C.~Urbach and I.~Wetzorke,
\newblock JHEP {\bf 09}, 071 (2005), [hep-lat/0507010].
%%CITATION = HEP-LAT 0507010;%%

\bibitem{Jagels:1994a}
C.~F. Jagels and L.~Reichel,
\newblock Numer. Linear Algebra Appl. {\bf 1(6)}, 555 (1994).

\bibitem{Saad:1993a}
Y.~Saad,
\newblock SIAM J. Sci. Comput. {\bf 14 (2)} (1993).

\bibitem{Glassner:1996gz}
U.~Gl{\"a}ssner {\em et~al.},
\newblock hep-lat/9605008.
%%CITATION = HEP-LAT 9605008;%%

\bibitem{Jegerlehner:1996pm}
B.~Jegerlehner,
\newblock hep-lat/9612014.
%%CITATION = HEP-LAT 9612014;%%

\bibitem{Jegerlehner:2003qp}
F.~Jegerlehner,
\newblock Nucl. Phys. Proc. Suppl. {\bf 126}, 325 (2004), [hep-ph/0310234].
%%CITATION = HEP-PH 0310234;%%

\bibitem{Edwards:1998wx}
R.~G. Edwards, U.~M. Heller and R.~Narayanan,
\newblock Phys. Rev. {\bf D59}, 094510 (1999), [hep-lat/9811030].
%%CITATION = HEP-LAT 9811030;%%

\bibitem{Giusti:2002sm}
L.~Giusti, C.~Hoelbling, M.~L{\"u}scher and H.~Wittig,
\newblock Comput. Phys. Commun. {\bf 153}, 31 (2003), [hep-lat/0212012].
%%CITATION = HEP-LAT 0212012;%%

\bibitem{Farchioni:2004us}
F.~Farchioni {\em et~al.},
\newblock Eur. Phys. J. {\bf C39}, 421 (2005), [hep-lat/0406039].
%%CITATION = HEP-LAT 0406039;%%

\bibitem{Jansen:1997yt}
K.~Jansen and C.~Liu,
\newblock Comput. Phys. Commun. {\bf 99}, 221 (1997), [hep-lat/9603008].
%%CITATION = HEP-LAT 9603008;%%

\bibitem{Burrage:1998a}
K.~Burrage and J.~Erhel,
\newblock Num. Lin. Alg. with Appl. {\bf 5}, 101 (1998).

\bibitem{Morgan:2002a}
R.~B. Morgan,
\newblock SIAM J. Sci. Comput. {\bf 24}, 20 (2002).

\bibitem{Sonneveld:1989cgs}
P.~Sonneveld,
\newblock SIAM J. Sci. Stat. Comput. {\bf 10}, 36 (1989).

\bibitem{Clark:2005sq}
M.~A. Clark, P.~de~Forcrand and A.~D. Kennedy,
\newblock PoS {\bf LAT2005}, 115 (2005), [hep-lat/0510004].
%%CITATION = HEP-LAT 0510004;%%

\bibitem{Clark:2006fx}
M.~A. Clark and A.~D. Kennedy,
\newblock hep-lat/0608015.
%%CITATION = HEP-LAT 0608015;%%

\bibitem{Fischer:1996th}
S.~Fischer {\em et~al.},
\newblock Comp. Phys. Commun. {\bf 98}, 20 (1996), [hep-lat/9602019].
%%CITATION = HEP-LAT 9602019;%%

\bibitem{Frommer:1994vn}
A.~Frommer, V.~Hannemann, B.~Nockel, T.~Lippert and K.~Schilling,
\newblock Int. J. Mod. Phys. {\bf C5}, 1073 (1994), [hep-lat/9404013].
%%CITATION = HEP-LAT 9404013;%%

\bibitem{Kalkreuter:1995mm}
T.~Kalkreuter and H.~Simma,
\newblock Comput. Phys. Commun. {\bf 93}, 33 (1996), [hep-lat/9507023].
%%CITATION = HEP-LAT 9507023;%%

\bibitem{Cundy:2002hv}
N.~Cundy, M.~Teper and U.~Wenger,
\newblock Phys. Rev. {\bf D66}, 094505 (2002), [hep-lat/0203030].
%%CITATION = HEP-LAT 0203030;%%

\bibitem{Sleijpen:1996aa}
G.~L.~G. Sleijpen and H.~A.~V. der Vorst,
\newblock SIAM Journal on Matrix Analysis and Applications {\bf 17}, 401
  (1996).

\bibitem{Fokkema:1998aa}
D.~R. Fokkema, G.~L.~G. Sleijpen and H.~A. Van~der Vorst,
\newblock J. Sci. Comput. {\bf 20}, 94 (1998).

\bibitem{Geus:2002}
R.~Geus,
\newblock {\em The Jacobi-Davidson algorithm for solving large sparse symmetric
  eigenvalue problems with application to the design of accelerator cavities},
\newblock PhD thesis, Swiss Federal Institute Of Technology Z{\"u}rich, 2002.

\end{thebibliography}
%%%%%%%%%%%%%%%%%%%%%%%%%%%%%%%%%%%%%%%%%%%%%%%%%%%%%%%%%%%%%%%%%%%%%%
%%%%%%%%%%%%%%%%%%%%%%%%%%%%%%%%%%%%%%%%%%%%%%%%%%%%%%%%%%%%%%%%%%%%%%

\end{document}